\documentclass[preprint]{aastex63}

\received{January 19, 2021}
\revised{March 31, 2021}
\accepted{April 8, 2021}

\shorttitle{Critical Collision Velocity for Water Ice}
\shortauthors{Hasegawa et al.}

\begin{document}

\title{Collisional Growth and Fragmentation of Dust Aggregates with Low Mass Ratios. I: Critical Collision Velocity for Water Ice}

\correspondingauthor{Yukihiko Hasegawa}
\email{yukih@ea.c.u-tokyo.ac.jp}

\author[0000-0001-7298-2144]{Yukihiko Hasegawa}
\affiliation{Graduate School of Arts \& Sciences, The University of Tokyo, 3-8-1, Komaba, Meguro, Tokyo 153-8902, Japan}

\author[0000-0001-9734-9601]{Takeru K. Suzuki}
\affiliation{Graduate School of Arts \& Sciences, The University of Tokyo, 3-8-1, Komaba, Meguro, Tokyo 153-8902, Japan}

\author[0000-0001-9659-658X]{Hidekazu Tanaka}
\affiliation{Astronomical Institute, Graduate School of Science, Tohoku University, 6-3, Aramaki, Aoba-ku, Sendai 980-8578, Japan}

\author[0000-0001-8808-2132]{Hiroshi Kobayashi}
\affiliation{Department of Physics, Nagoya University, Nagoya, Aichi 464-8602, Japan}

\author[0000-0002-6710-1768]{Koji Wada}
\affiliation{Planetary Exploration Research Center, Chiba Institute of Technology, 2-17-1, Tsudanuma, Narashino, Chiba 275-0016, Japan}





\begin{abstract}

We investigated fundamental processes of collisional sticking and fragmentation of dust aggregates by carrying out $N$-body simulations of submicron-sized icy dust monomers.
We examined the condition for collisional growth of two colliding dust aggregates in a wide range of the mass ratio, 1-64.
We found that the mass transfer from a larger dust aggregate to a smaller one is a dominant process in collisions with a mass ratio of 2-30 and impact velocity of $\approx$ 30-170 $\mathrm{m ~ s^{-1}}$.
As a result, the critical velocity, $v_{\mathrm{fra}}$, for fragmentation of the largest body is considerably reduced for such unequal-mass collisions; $v_{\mathrm{fra}}$ of collisions with a mass ratio of 3 is about half of that obtained from equal-mass collisions.
The impact velocity is generally higher for collisions between dust aggregates with higher mass ratios because of the difference between the radial drift velocities in the typical condition of protoplanetary disks.
Therefore, the reduced $v_{\mathrm{fra}}$ for unequal-mass collisions would delay growth of dust grains in the inner region of protoplanetary disks.

\end{abstract}




\section{Introduction} \label{sec:int}

In protoplanetary disks (hereafter PPDs) dust grains grow to larger bodies mainly though collisional sticking.
As a result of the growth of dust grains, their size and internal structure change, which affects the dynamics of dust grains in the gas component of PPDs (Adachi et al. 1976; Weidenschilling 1977; Nakagawa et al. 1986; Youdin \& Shu 2002; Michikoshi \& Inutsuka 2006; Brauer et al. 2008; Bai \& Stone 2010; Birnstiel et al. 2010; Okuzumi et al. 2012), observed properties of PPDs (Dullemond \& Dominik 2005; Tanaka et al. 2005; Kataoka et al. 2015, 2016a, b; Stephens et al. 2017; Okuzumi \& Tazaki 2019; Tazaki et al. 2019a, b) and the subsequent planetesimal and/or planet formation (Nakagawa et al. 1981, 1986; Okuzumi et al. 2012; Kataoka et al. 2013; Dr\c{a}\.{z}kowska \& Dullemond 2014; Krijt et al. 2015).
It is generally considered that dust grains are the aggregates of (sub-)\micron-sized, small, dense and spherical particles, which are called monomers.
When dust aggregates grow, however, there are two major obstacles that prevent further growth, i.e., the radial drift barrier (Adachi et al. 1976; Weidenschilling 1977; Brauer et al. 2008; Okuzumi et al. 2012; Kataoka et al. 2013; Arakawa \& Nakamoto 2016) and the fragmentation barrier (Dominik \& Tielens 1997; Blum \& Wurm 2000; Brauer et al. 2008; Wada et al. 2009, 2013; Birnstiel et al. 2010; Krijt et al. 2015).

In typical PPD conditions, the gas drag force that acts on dust aggregates transfers the angular momentum of the solid component to the gas component, and then solid bodies suffer from rapid inward drift to the central star.
In the minimum-mass solar nebula (MMSN) model (Hayashi 1981; Hayashi et al. 1985), which is a standard model of PPDs, the timescale of the radial drift is $\sim 10^2$ years for meter-sized boulders located at 1 au from the central star (Adachi et al. 1976; Brauer et al. 2008).
Since this timescale is estimated to be shorter than the growth timescale for spherical-shaped compact dust grains, solid particles are expected to fall onto the central star before growing into km-sized planetesimals (Brauer et al. 2008).
On the other hand, it is suggested that fluffy dust aggregates possibly overcome the radial drift barrier because the aerodynamical property is altered (Okuzumi et al. 2012).
The gas drag force of fluffy dust aggregates changes from the Epstein regime to the Stokes regime at the earlier epoch of the dust growth (i.e., before the encounter with the drift barrier).
At the Stokes regime, the collision frequency increases with the dust size.
This acceleration in collisional growth enables dust aggregates to overcome the radial drift barrier.

The collision velocity between dust aggregates increases with the dust size (Krijt et al. 2015).
However, when the collision velocity is high, collisional fragmentation is more dominant than growth by collisional sticking.
In the case when the collisional fragmentation delays the dust growth even if does not hinder, the radial drift barrier may become severe and prevent dust aggregates from further growing to planetesimals.
The minimum collision velocity for the fragmentation of dust aggregates, i.e., the critical collision velocity for fragmentation, depends on material properties of the monomers.
There are many previous studies of both laboratory experiments and numerical simulations for collisions between dust aggregates and/or solids (Dominik \& Tielens 1997; Blum \& Wurm 2000; Wada et al. 2007, 2008, 2009, 2013; Suyama et al. 2008, 2012; Teiser \& Wurm 2009; Kobayashi \& Tanaka 2010; Gundlach et al. 2011; Ringl et al. 2012; Meru et al. 2013; Seizinger et al. 2013; Gundlach \& Blum 2015; Schr\"{a}pler et al. 2018).
G\"{u}ttler et al. (2010) reviewed results of previous laboratory experiments and showed that the critical collision velocity for fragmentation of dust aggregates composed of micron-sized silicate monomers is $\sim 1 ~ \mathrm{m ~ s^{-1}}$.
Teiser \& Wurm (2009) carried out laboratory experiments using micron-sized monomers and showed that silicate targets much larger than silicate projectiles can accrete the projectiles in central collisions even at 56.5 $\mathrm{m ~ s^{-1}}$.
Gundlach \& Blum (2015) performed laboratory experiments on collisions 
\replaced{between dust aggregates composed of micron-sized water-ice monomers}{
of micron-sized water-ice particles with dust aggregates grown on the cold plate
}
 and showed that the collisional sticking occurs at the collision velocity lower than 9.6 $\mathrm{m ~ s^{-1}}$.
Their results are in agreement with the critical velocity estimated in Wada et al. (2013).

Since it is quite difficult to track growth from small dust grains to planetesimals by direct numerical simulations at the moment, analytical approaches have been adopted (Brauer et al. 2008; Birnstiel et al. 2010; Okuzumi et al. 2012; Dr\c{a}\.{z}kowska \& Dullemond 2014; Krijt et al. 2015).
These works incorporate results and recipes obtained from numerical simulations (Wada et al. 2009, 2013; Suyama et al. 2012; Seizinger et al. 2013) to construct analytical models.
As examples of simulation studies, Suyama et al. (2012) carried out $N$-body simulations of sequential head-on collisions between different-sized dust aggregates with low collision velocity and improved their model of collisional compression.
Wada et al. (2013) carried out simulations of offset collisions between dust aggregates with different sizes and obtained the critical collisional fragmentation velocity.
Seizinger et al. (2013) performed simulations of collisions between targets composed of about $10^5$ monomers and various small colliding bodies and showed that the critical collisional fragmentation velocity depends on the projectile mass.
However, numerical simulations of collisions with mass ratios of 1-10 have not been performed so far.
Wada et al. (2009) carried out only equal-mass collisions between dust aggregates with the total monomer number of $\approx 1.6 \times 10^4$ at most.
Wada et al. (2013) investigated not only equal-mass but also unequal-mass collisions, but they carried out simulations of the collisions with mass ratios of 16 and 64 only.
Suyama et al. (2012) performed simulations of collisions between icy dust aggregates with a mass ratio of 4, but the collision velocity is lower than 5 $\mathrm{m ~ s^{-1}}$ since they focused on collisional compression of dust aggregates and did not discuss the collisional fragmentation.

The purpose of the present paper is to study basic physical properties of collisions between water-ice dust aggregates with various mass ratios; we particularly focus on unequal-mass collisions that have not been covered by previous works.
In section \ref{sec:mod}, we briefly describe our simulation method.
In section \ref{sec:res}, results of our simulations are shown.
Using these results, we derive the condition for the collisional growth of dust aggregates.
In section \ref{sec:dis}, we discuss the collision velocity in PPDs and the mass of small fragments reproduced through collisions.
Summary of this study is in section \ref{sec:sum}.

\section{Simulation model} \label{sec:mod}

We perform three-dimensional numerical simulations of collisions between two water-ice dust aggregates.
The numerical $N$-body code is the same as one used in Wada et al. (2009).
Here we briefly explain the outline, whereas readers may refer to Wada et al. (2007) for the details.
Dust monomers are treated as elastic spheres.
When two elastic spheres contact with each other, the interaction between spheres is given in the JKR theory (Johnson et al. 1971; Johnson 1987).
Four motions of the two contact particles, i.e., compression/adhesion, rolling, sliding, and twisting, are taken into account (Dominik \& Tielens 1997; Wada et al. 2007; Seizinger et al. 2013).
Only compression/adhesion is the normal motion to the contact surface, and the other components are tangential motions.
The tangential motions are classified into (i) the elastic regime without energy dissipation because of small displacements and (ii) the inelastic regime with energy dissipation originating from the irregularities on the monomer surfaces.
We numerically simulate the motions of dust monomers constituting dust aggregates.

We assume spherical dust monomers made of water ice, with the same radius of 0.1 \micron ~ and an internal density of 1 $\mathrm{g ~ cm^{-3}}$ (Wada et al. 2007).
For properties of ice, the surface energy is $100 ~ \mathrm{mJ ~ m^{-2}} = 100 ~ \mathrm{erg ~ cm^{-2}}$, Young's modulus is $7 ~ \mathrm{GPa} = 7 \times 10^{10}$ $\mathrm{dyn ~ cm^{-2}}$, and Poisson's ratio is 0.25 (Wada et al. 2007).
In this paper, we adopt 8 \r{A} as the critical rolling displacement (Wada et al. 2009).
We assume that the initial dust aggregates before collisions are ballistic particle-cluster aggregation (BPCA) clusters formed by sequential hit-and-stick collisions between a cluster (aggregate) and multiple single particles (monomers) (Wada et al. 2013).

\begin{figure}[t]
  \plotone{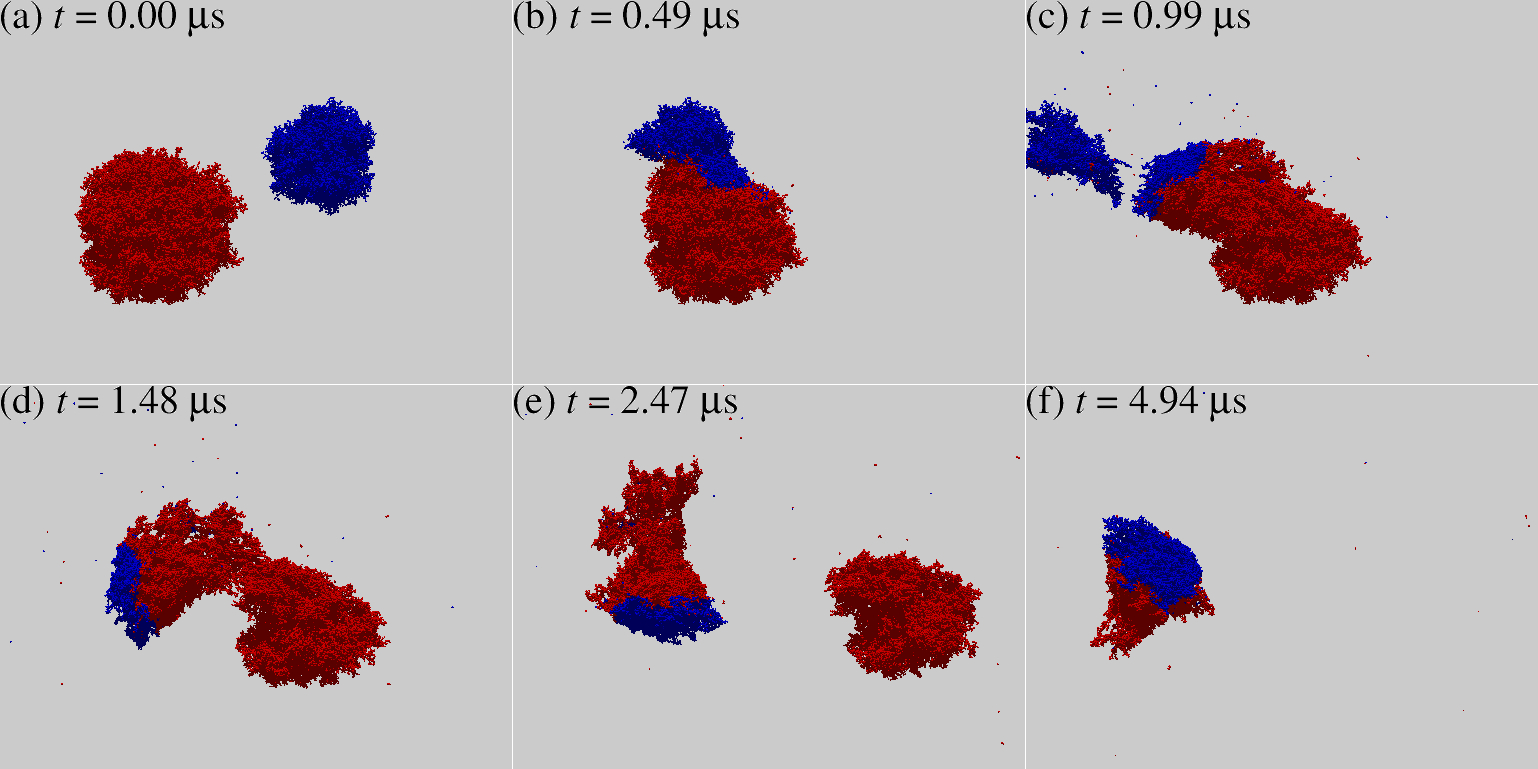}
  \caption{Snapshots of an offset collision with $b_{\mathrm{off}} / b_{\mathrm{max}} = 0.55$ between a target of $N_{\mathrm{tar}} = 65536$ and a projectile of $N_{\mathrm{pro}} = 21845$ (i.e., $N_{\mathrm{tar}} / N_{\mathrm{pro}} = 3$) at $v_{\mathrm{col}} = 44 ~ \mathrm{m ~ s^{-1}}$. Red (blue) indicates monomers that belonged to the target (projectile).}
  \label{fig:snapshot_4_45_1_zx}
\end{figure}

The target is defined as the larger colliding dust aggregates, and the smaller collider is called the projectile.
Figure \ref{fig:snapshot_4_45_1_zx} shows an example of an offset collision between a target of $N_{\mathrm{tar}} = 65536$ and a projectile of $N_{\mathrm{pro}} = 21845$, where $N_{\mathrm{tar}}$ is the number of monomers in the target and $N_{\mathrm{pro}}$ is that in the projectile.
We note that the ratio of monomer numbers, $N_{\mathrm{tar}} / N_{\mathrm{pro}}$, corresponds to the mass ratio because of equal-sized monomers.
We focus on the collisional outcomes such as $N_{\mathrm{lar}}$, the number of monomers in the largest dust aggregate resulting from the collision.
The simulation shown in Figure \ref{fig:snapshot_4_45_1_zx} results in the $N_{\mathrm{lar}} = 42285$ [the right dust aggregate in Figure \ref{fig:snapshot_4_45_1_zx}-(e)].

We perform four simulation runs with randomly generated targets and projectiles for a set of the parameters, the number of monomers of the target, $N_{\mathrm{tar}}$, and the projectile, $N_{\mathrm{pro}}$, the collision velocity, $v_{\mathrm{col}}$, and the impact parameter, $b_{\mathrm{off}}$.
For quantitative analysis we take the average of the four runs with the same set of $N_{\mathrm{tar}}$, $N_{\mathrm{pro}}$, $v_{\mathrm{col}}$, and $b_{\mathrm{off}}$.
We represent the average of a variable $A$ taken over the four runs as $\bar{A}$.
The input parameters and the collisional outcomes of our simulations, and the averages of variables are summarized in Table \ref{tab:variables}.
\begin{deluxetable}{ll}
  \tablecaption{Definition of symbols used in this paper.}
  \label{tab:variables}
  \tablehead
  {
    \colhead{Symbol} & \colhead{Definition}
  }
  \startdata
      \multicolumn{2}{l}{Input parameters} \\
      \hline
      $N_{\mathrm{tar}}$ & Number of monomers constituting the target \\
      $N_{\mathrm{pro}}$ & Number of monomers constituting the projectile \\
      $v_{\mathrm{col}}$ & Collision velocity between the target and the projectile \\
      $b_{\mathrm{off}}$ & Impact parameter \\
      \hline
      \multicolumn{2}{l}{Collisional outcomes} \\
      \hline
      $N_{\mathrm{lar}}$ & Number of monomers in the largest remnant \\
      $N_{\mathrm{2nd}}$ & Number of monomers in the second largest remnant \\
      $N_{\mathrm{3rd}}$ & Number of monomers in the third largest remnant \\
      $f_{\mathrm{gro}}$ & Collisional growth efficiency of the monomer number of the largest remnant [Equation (\ref{equ:growth_efficiency_1st_def})] \\
      $f_{\mathrm{2nd}}$ & Collisional growth efficiency of the monomer number of the second largest remnant [Equation (\ref{equ:growth_efficiency_2nd_def})] \\
      $v_{\mathrm{fra}}$ & Lower limit of the collision velocity on collisional fragmentation, called the critical collisional fragmentation velocity \\
      \hline
      \multicolumn{2}{l}{Averages} \\
      \hline
      $\bar{A}$ & Average of a variable $A$ taken over the four runs \\
      $\langle A \rangle$ & Average of a variable $A$ weighted over the impact parameter, called the $b_{\mathrm{off}}$-weighted average [Equation (\ref{equ:b_weight_def})] \\
  \enddata
\end{deluxetable}
In this study, 33280 runs were carried out 
\added{
(Appendix \ref{sec:app_A}).
Numerical results of our simulations will be further analyzed in our next paper.
We plan to release the numerical data when the next paper is published.
}

\section{Results} \label{sec:res}

\subsection{Growth efficiency of the largest remnant} \label{sec:res_lar}

As shown in Figure \ref{fig:snapshot_4_45_1_zx}, a collision changes the number of monomers in dust aggregates.
If the largest remnant is larger than the target, the collision induces the collisional growth of the target.
We investigate the collisional growth efficiency as
\begin{equation}
  f_{\mathrm{gro}} \equiv \frac{N_{\mathrm{lar}} - N_{\mathrm{tar}}}{N_{\mathrm{pro}}}
  \mathrm{.}
  \label{equ:growth_efficiency_1st_def}
\end{equation}
In addition, considering the frequency dependence on the impact parameter and we take the average of a variable $A$ weighted over the impact parameter,
\begin{equation}
  \langle A \rangle \equiv \left . \int _0^{b_{\mathrm{max}}} 2 \pi b_{\mathrm{off}} \bar{A} \mathrm{d}b_{\mathrm{off}} ~ \right / \int _0^{b_{\mathrm{max}}} 2 \pi b_{\mathrm{off}} \mathrm{d}b_{\mathrm{off}}
  \mathrm{,}
  \label{equ:b_weight_def}
\end{equation}
which is called the $b_{\mathrm{off}}$-weighted average, where $b_{\mathrm{max}}$ is the maximum value of the impact parameter.
The maximum impact parameter is the sum of the characteristic radii, $(5/3)^{1/2} r_{\mathrm{gyr}}$, of the target and the projectile, where $r_{\mathrm{gyr}}$ is the gyration radius of the dust aggregate (Mukai et al. 1992; Wada et al. 2013).
The lower limit of the collision velocity on collisional fragmentation, $v_{\mathrm{fra}}$, is defined as the collision velocity with $\langle N_{\mathrm{lar}} \rangle = N_{\mathrm{tar}}$ and is hereafter called the critical collisional fragmentation velocity, which is equivalent to the critical collision velocity for disruption of dust aggregates in Wada et al. (2009).

\begin{figure}[t]
  \plottwo{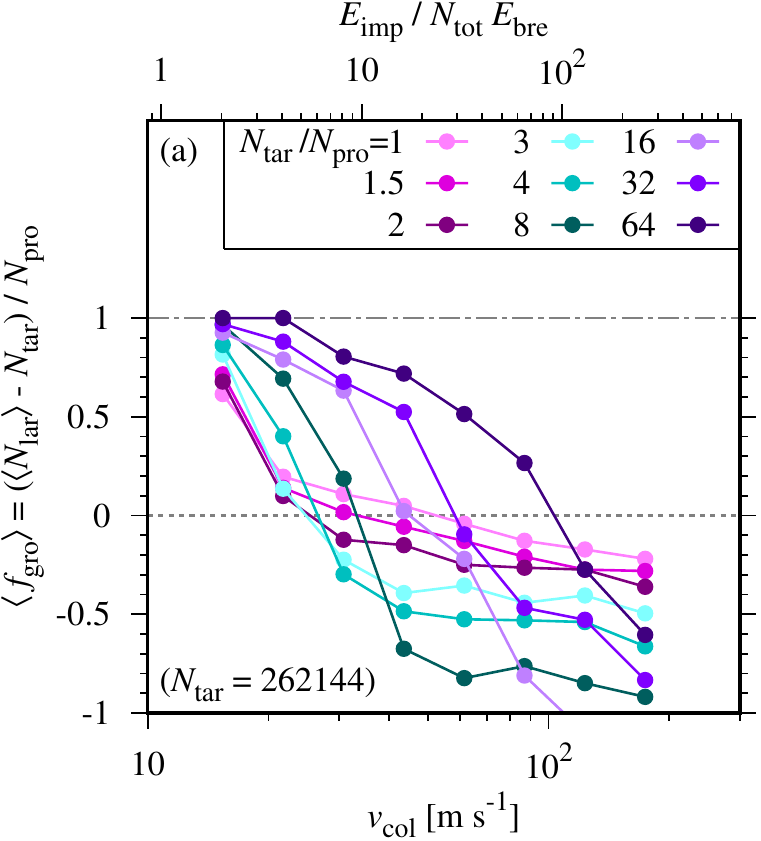}{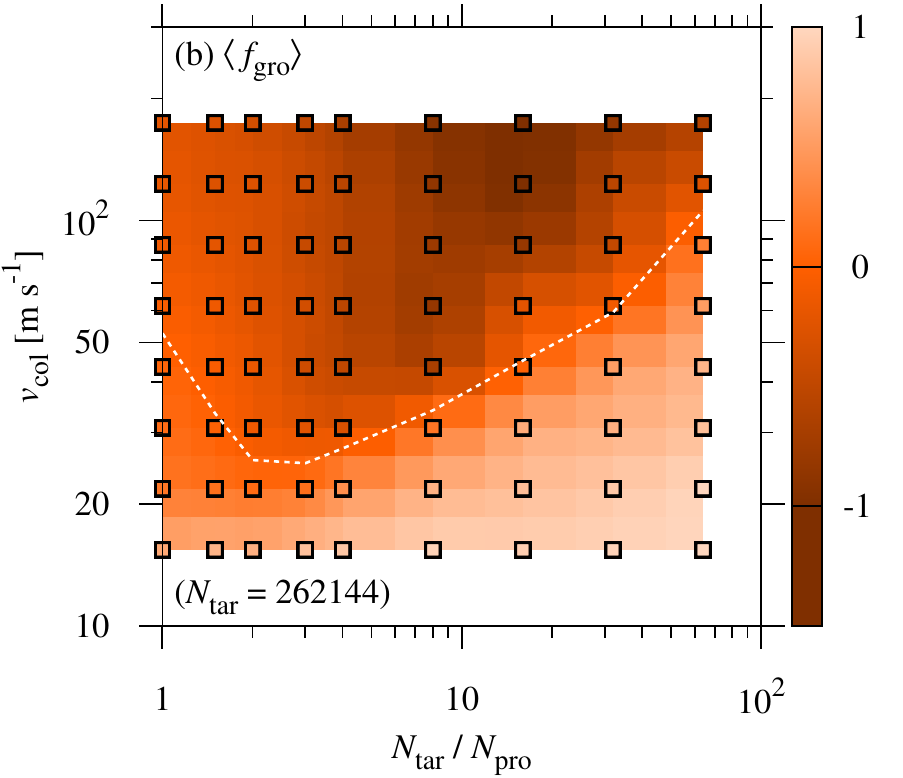}
  \caption{(a) Growth efficiency of the $b_{\mathrm{off}}$-weighted average monomer number of the largest remnant, $\langle f_{\mathrm{gro}} \rangle$, against the collision velocity $v_{\mathrm{col}}$ for $N_{\mathrm{tar}} = 262144$ (filled circles and solid lines). Colors represent the mass ratios of colliders, $N_{\mathrm{tar}} / N_{\mathrm{pro}}$. The dotted line marks $\langle f_{\mathrm{gro}} \rangle = 0$, and the intersections between solid lines and the dotted line correspond to the critical collisional fragmentation velocity, $v_{\mathrm{fra}}$. The dash-dotted line marks the upper limit of the growth efficiency, $\langle f_{\mathrm{gro}} \rangle = 1$, equivalent to $\langle N_{\mathrm{lar}} \rangle = N_{\mathrm{tar}} + N_{\mathrm{pro}}$. The scale of the upper horizontal axis is given by $E_{\mathrm{imp}} / N_{\mathrm{tot}} E_{\mathrm{bre}}$, where $E_{\mathrm{imp}} = (1/2) N_{\mathrm{tot}} m_{\mathrm{mon}} (v_{\mathrm{col}} / 2)^2$ is the impact energy, $N_{\mathrm{tot}} = N_{\mathrm{tar}} + N_{\mathrm{pro}}$ is the sum of the monomer numbers constituting the target and the projectile, $m_{\mathrm{mon}}$ is the monomer mass, and $E_{\mathrm{bre}}$ is the energy for breaking a single contact between two identical particles. (b) Growth efficiency $\langle f_{\mathrm{gro}} \rangle$ on the $N_{\mathrm{tar}} / N_{\mathrm{pro}}$-$v_{\mathrm{col}}$ plane for $N_{\mathrm{tar}} = 262144$ (colors and squares). The squares represent the sets of our simulations. The white dotted line marks $\langle f_{\mathrm{gro}} \rangle = 0$, i.e., $v_{\mathrm{col}} = v_{\mathrm{fra}}$.}
  \label{fig:Ntop_vcol_f1st_Ntar262144}
\end{figure}

Figure \ref{fig:Ntop_vcol_f1st_Ntar262144} shows $\langle f_{\mathrm{gro}} \rangle$ against the collision velocity $v_{\mathrm{col}}$ for different initial ratios of monomer numbers, $N_{\mathrm{tar}} / N_{\mathrm{pro}}$, for $N_{\mathrm{tar}} = 262144$.
The range of $\langle f_{\mathrm{gro}} \rangle$ is given by $(1 - N_{\mathrm{tar}}) / N_{\mathrm{pro}} ~ (\approx - N_{\mathrm{tar}} / N_{\mathrm{pro}}) \le \langle f_{\mathrm{gro}} \rangle \le 1$ because $1 \le \langle N_{\mathrm{lar}} \rangle \le N_{\mathrm{tar}} + N_{\mathrm{pro}}$.
The intersections between solid lines and the dotted line $\langle f_{\mathrm{gro}} \rangle = 0$ correspond to $v_{\mathrm{fra}}$ for each $N_{\mathrm{tar}} / N_{\mathrm{pro}}$.
The horizontal axis and vertical axis of Figure \ref{fig:Ntop_vcol_f1st_Ntar262144}-(b) are given by $N_{\mathrm{tar}} / N_{\mathrm{pro}}$ and $v_{\mathrm{col}}$, respectively, and the dotted line marks $v_{\mathrm{col}} = v_{\mathrm{fra}}$.
All results for other $N_{\mathrm{tar}}$ and/or $N_{\mathrm{pro}}$, and a comparison with results obtained by Wada et al. (2013) are shown in Appendix \ref{sec:app_B}.

For $1.5 \le N_{\mathrm{tar}} / N_{\mathrm{pro}} \le 8$, which have not been studied in previous works, one can clearly see that $v_{\mathrm{fra}} \approx 30 ~ \mathrm{m ~ s^{-1}}$, which is much lower than the critical collision velocities, $=$ 67-90 $\mathrm{m ~ s^{-1}}$, obtained from the simulations with $N_{\mathrm{tar}} / N_{\mathrm{pro}} =$ 1, 16, and 64 in Wada et al. (2013).
Since these values are also lower than the maximum radial drift speed $= 54 ~ \mathrm{m ~ s^{-1}}$ (Nakagawa et al. 1986; Okuzumi et al. 2012), these unequal-mass collisions could severely affect sticking growth of dust aggregates.

For equal-mass collisions, the simulations with $N_{\mathrm{tot}} \equiv N_{\mathrm{tar}} + N_{\mathrm{pro}} = 524288$ result in $v_{\mathrm{fra}} = 53 ~ \mathrm{m ~ s^{-1}}$.
In contrast, Wada et al. (2009) reported slightly higher $v_{\mathrm{fra}} =$ 57-68 $\mathrm{m ~ s^{-1}}$ from their simulations with the smaller number of monomers, $N_{\mathrm{tot}} =$ 1000, 4000 and 16000.
Difference between this paper and Wada et al. (2009) is in the number of monomers.
We found that efficient fragmentation occurs for offset collisions with $b_{\mathrm{off}} / b_{\mathrm{max}} \gtrsim 0.5$ in our simulations with larger $N_{\mathrm{tot}}$ (Appendix \ref{sec:app_C}).
These offset collisions largely contribute to $b_{\mathrm{off}}$-weighted average because of the areal fraction [Equation (\ref{equ:b_weight_def})], which reduces $v_{\mathrm{fra}}$.

\begin{figure}[t]
  \plotone{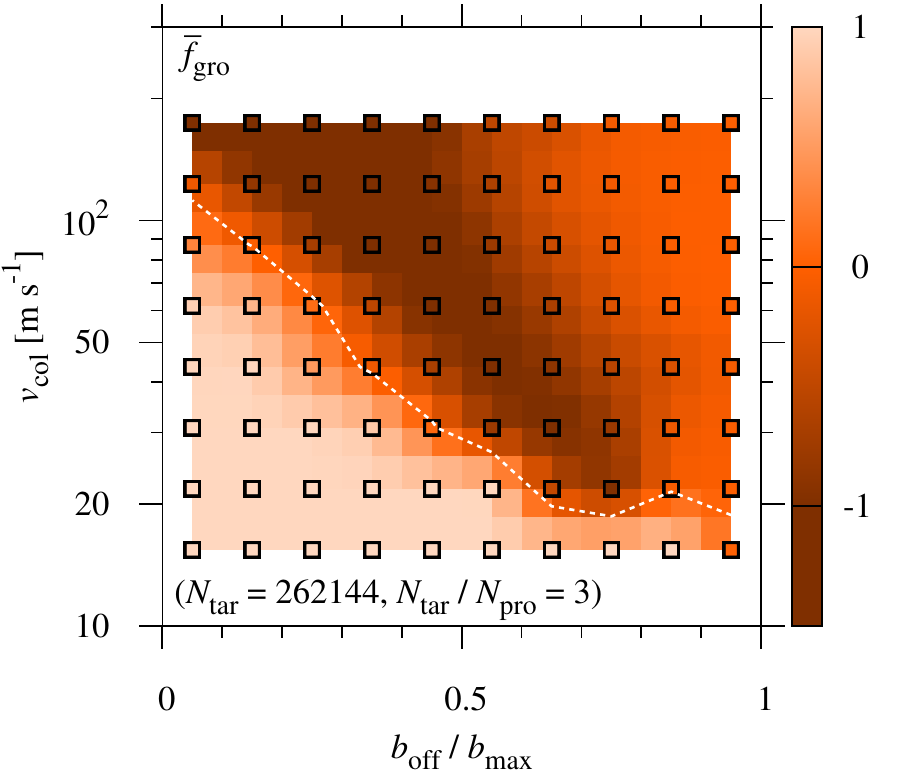}
  \caption{Growth efficiency ${\bar{f}}_{\mathrm{gro}}$ on the $b_{\mathrm{off}} / b_{\mathrm{max}}$-$v_{\mathrm{col}}$ plane for $N_{\mathrm{tar}} = 262144$ and $N_{\mathrm{tar}} / N_{\mathrm{pro}} = 3$ (colors and squares). The squares represent the sets of our simulations. The white dotted line marks ${\bar{f}}_{\mathrm{gro}} = 0$.}
  \label{fig:boff_vcol_f1st_Ntar262144_Ntop3}
\end{figure}

We examine detailed properties of unequal-mass collisions.
Here we focus on the cases with $N_{\mathrm{tar}} / N_{\mathrm{pro}} = 3$, which give the minimum $v_{\mathrm{fra}} = 25 ~ \mathrm{m ~ s^{-1}}$ for $N_{\mathrm{tar}} = 262144$ [Figure \ref{fig:Ntop_vcol_f1st_Ntar262144}-(b)].
In order to see the dependence on the impact parameter, we present the collisional growth efficiency, ${\bar{f}}_{\mathrm{gro}} \equiv ( {\bar{N}}_{\mathrm{lar}} - N_{\mathrm{tar}} ) / N_{\mathrm{pro}}$, without averaged over $b_{\mathrm{off}}$ in Figure \ref{fig:boff_vcol_f1st_Ntar262144_Ntop3}.
We again note that the range of ${\bar{f}}_{\mathrm{gro}}$ is given by $- N_{\mathrm{tar}} / N_{\mathrm{pro}} \lesssim {\bar{f}}_{\mathrm{gro}} \le 1$ [see Equation (\ref{equ:growth_efficiency_1st_def})].

Figure \ref{fig:boff_vcol_f1st_Ntar262144_Ntop3} exhibits that ${\bar{f}}_{\mathrm{gro}} < 0$ for offset collision with $b_{\mathrm{off}} / b_{\mathrm{max}} \gtrsim 0.5$, even at the quite low collision velocity, $v_{\mathrm{col}} \gtrsim 20 ~ \mathrm{m ~ s^{-1}}$.
As a result, the collision velocity that gives ${\bar{f}}_{\mathrm{gro}} = 0$ for these offset collisions is $\approx 20 ~ \mathrm{m ~ s^{-1}}$, which is the main reason of the low critical collisional fragmentation velocity of $v_{\mathrm{fra}} = 25 ~ \mathrm{m ~ s^{-1}}$.
This tendency also holds for other unequal-mass collisions with $N_{\mathrm{tar}} / N_{\mathrm{pro}} =$ 1.5, 2, 4, 8 and 16.
Therefore, relatively low $v_{\mathrm{fra}}$ is obtained for $1 < N_{\mathrm{tar}} / N_{\mathrm{pro}} \lesssim 20$.

\subsection{Mass transfer from target to projectile} \label{sec:res_2nd}

The collisions with $N_{\mathrm{tar}} / N_{\mathrm{pro}} \lesssim 20$ tend to give large $N_{\mathrm{2nd}} / N_{\mathrm{lar}}$, where $N_{\mathrm{2nd}}$ is the number of monomers in the second largest remnant.
The simulation for $N_{\mathrm{tar}} / N_{\mathrm{pro}} = 3$ and $b_{\mathrm{off}} / b_{\mathrm{max}} = 0.55$ shown in Figure \ref{fig:snapshot_4_45_1_zx} results in $N_{\mathrm{lar}} = 42285$ [the right dust aggregate in Figure \ref{fig:snapshot_4_45_1_zx}-(e)] and $N_{\mathrm{2nd}} = 36211$ [the left dust aggregate in Figure \ref{fig:snapshot_4_45_1_zx}-(f)].
The second remnant seen in Figure \ref{fig:snapshot_4_45_1_zx}-(f) consists of not only the projectile but also a part of the target; the offset collision triggers the erosion with mass transfer from the target to the projectile.
As a result, we obtained $N_{\mathrm{2nd}} > N_{\mathrm{pro}}$ and $N_{\mathrm{lar}} < N_{\mathrm{tar}}$.
We note that the average of the four runs with the same parameter set also gives ${\bar{N}}_{\mathrm{lar}} = 4.2 \times 10^4 < N_{\mathrm{tar}}$ and ${\bar{N}}_{\mathrm{2nd}} = 3.6 \times 10^4 > N_{\mathrm{pro}}$; the transfer erosion is not an occasional event for this particular run but universal phenomena for this parameter set.
The erosion with mass transfer dominantly occurs for collisions with $v_{\mathrm{col}} = 44 ~ \mathrm{m ~ s^{-1}}$ and the initial mass ratio with $1 < N_{\mathrm{tar}} / N_{\mathrm{pro}} \le 8$, which is the main reason why $\langle N_{\mathrm{lar}} \rangle < N_{\mathrm{tar}}$ in this parameter range (Figure \ref{fig:Ntop_vcol_f1st_Ntar262144}).

\begin{figure}[t]
  \plotone{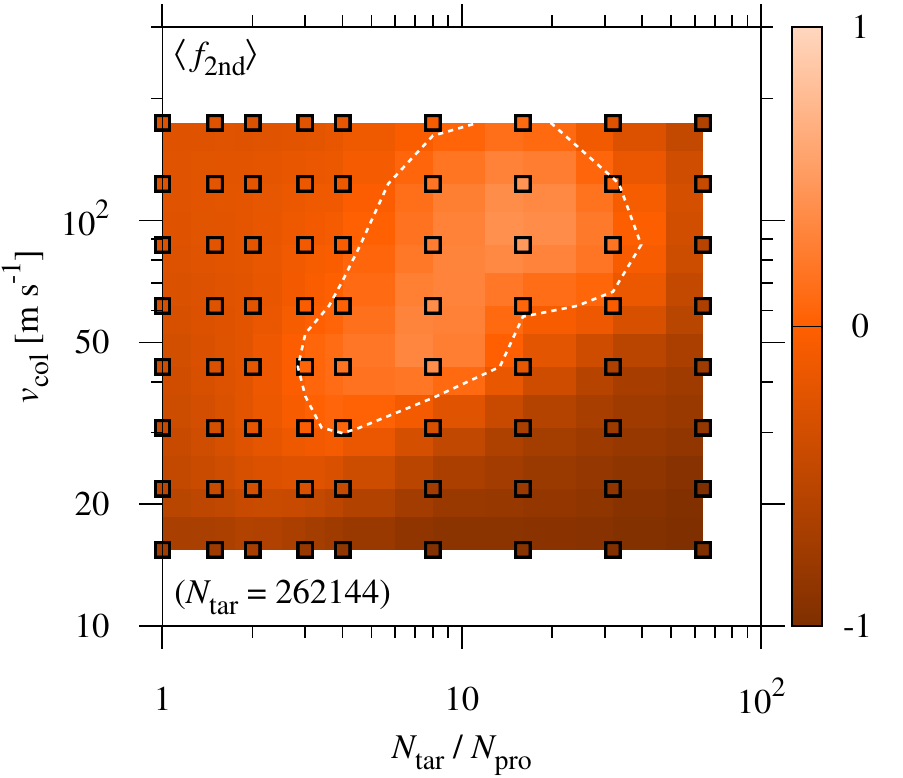}
  \caption{Growth efficiency of the $b_{\mathrm{off}}$-weighted average monomer number of the second remnant, $\langle f_{\mathrm{2nd}} \rangle$, on the $N_{\mathrm{tar}} / N_{\mathrm{pro}}$-$v_{\mathrm{col}}$ plane for $N_{\mathrm{tar}} = 262144$ (colors and squares). The squares are the same as Figure \ref{fig:Ntop_vcol_f1st_Ntar262144}. The white dotted line marks $\langle f_{\mathrm{2nd}} \rangle = 0$.}
  \label{fig:Ntop_vcol_f2nd_Ntar262144}
\end{figure}

Figure \ref{fig:Ntop_vcol_f2nd_Ntar262144} shows the $b_{\mathrm{off}}$-weighted collisional growth efficiency of the second remnant defined by
\begin{equation}
  \langle f_{\mathrm{2nd}} \rangle \equiv \frac{\langle N_{\mathrm{2nd}} \rangle - N_{\mathrm{pro}}}{N_{\mathrm{pro}}}
  \mathrm{,}
  \label{equ:growth_efficiency_2nd_def}
\end{equation}
against $N_{\mathrm{tar}} / N_{\mathrm{pro}}$ and $v_{\mathrm{col}}$ for $N_{\mathrm{tar}} = 262144$.
The range of $\langle f_{\mathrm{2nd}} \rangle$ is $-1 \le \langle f_{\mathrm{2nd}} \rangle \le [(N_{\mathrm{tar}} / N_{\mathrm{pro}}) - 1] / 2$.
The second largest remnant mass larger than the projectile, i.e., $\langle f_{\mathrm{2nd}} \rangle > 0$, appears for the cases of $N_{\mathrm{tar}} / N_{\mathrm{pro}} =$ 3-32.
For $N_{\mathrm{tar}} / N_{\mathrm{pro}} =$ 3-16 and $v_{\mathrm{col}} =$ 30-50 $\mathrm{m ~ s^{-1}}$, the projectile mass increases due to the mass transfer from the target (see Figure \ref{fig:snapshot_4_45_1_zx}), which achieves $\langle f_{\mathrm{2nd}} \rangle > 0$.
On the other hand, the second remnant would be formed from a fragment of the target for $N_{\mathrm{tar}} / N_{\mathrm{pro}} \gtrsim 8$ and $v_{\mathrm{col}} \gtrsim 10^2 ~ \mathrm{m ~ s^{-1}}$, while such a high collision velocity between dust aggregates may not be realized in PPDs.

All the cases with $N_{\mathrm{tar}} / N_{\mathrm{pro}} =$ 1.5 and 2 show negative $\langle f_{\mathrm{2nd}} \rangle$.
However, the mass transfer effectively occurs in some offset collisions.
For $N_{\mathrm{tar}} / N_{\mathrm{pro}} = 2$, ${\bar{f}}_{\mathrm{2nd}} \equiv ( {\bar{N}}_{\mathrm{2nd}} - N_{\mathrm{pro}} ) / N_{\mathrm{pro}}$ is positive for the collisions with $0.5 \lesssim b_{\mathrm{off}} / b_{\mathrm{max}} \lesssim 0.8$ and $20 ~ \mathrm{m ~ s^{-1}} \lesssim v_{\mathrm{col}} \lesssim 50 ~ \mathrm{m ~ s^{-1}}$, and the maximum ${\bar{f}}_{\mathrm{2nd}} = 0.23$ for $b_{\mathrm{off}} / b_{\mathrm{max}} = 0.65$ and $v_{\mathrm{col}} = 31 ~ \mathrm{m ~ s^{-1}}$.
For $N_{\mathrm{tar}} / N_{\mathrm{pro}} = 1.5$, the mass transfer increases the projectile mass to give $N_{\mathrm{2nd}} > N_{\mathrm{pro}}$ via grazing collisions with $b_{\mathrm{off}} / b_{\mathrm{max}} = 0.95$.
In summary, the mass transfer from the target to the projectile is the reason why the critical collisional fragmentation velocity of unequal-mass collisions is lower than that for equal-mass collisions.

\subsection{Small fragments} \label{sec:res_3rd}

\begin{figure}[t]
  \plottwo{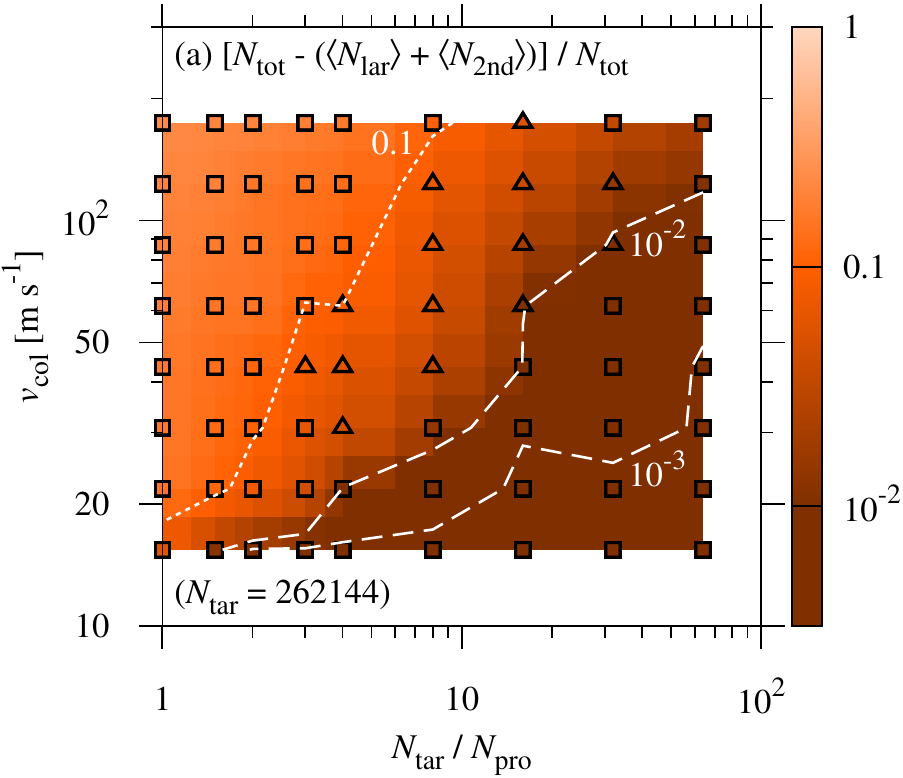}{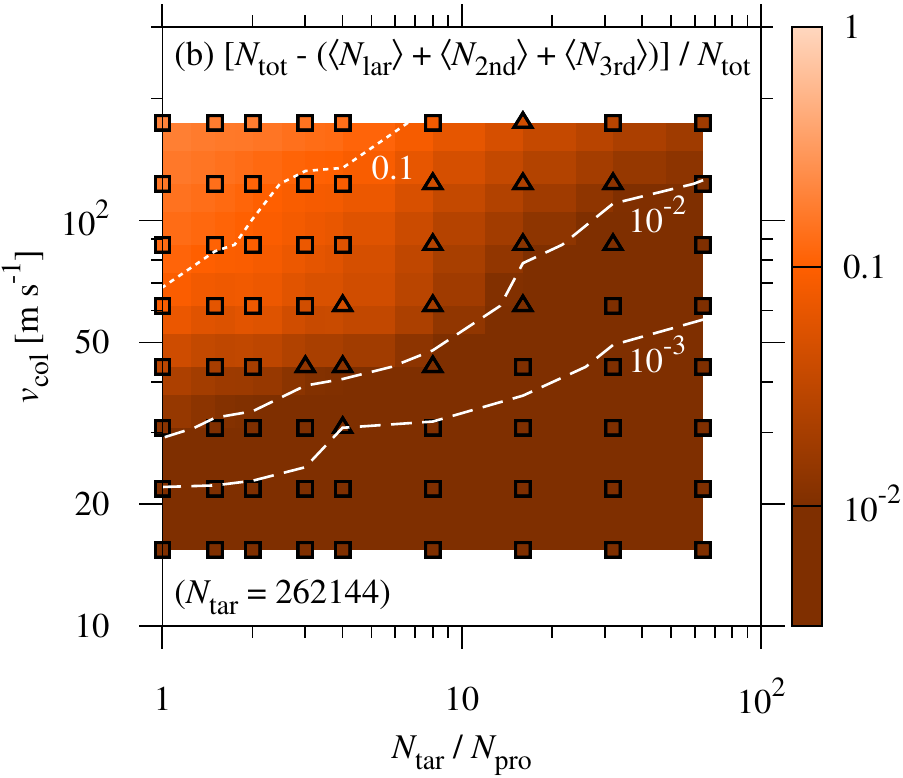}
  \caption{(a) Total monomer number of ejecta except for the two large remnants, normalized by the total monomer number, $[N_{\mathrm{tot}} - (\langle N_{\mathrm{lar}} \rangle + \langle N_{\mathrm{2nd}} \rangle )] / N_{\mathrm{tot}}$, on the $N_{\mathrm{tar}} / N_{\mathrm{pro}}$-$v_{\mathrm{col}}$ plane for $N_{\mathrm{tar}} = 262144$ (color contour). Squares (triangles) indicates sets of the simulations that give $\langle f_{\mathrm{2nd}} \rangle < (>) ~ 0$. The white dotted line marks $\langle N_{\mathrm{lar}} \rangle + \langle N_{\mathrm{2nd}} \rangle = 0.9 N_{\mathrm{tot}}$. The white dashed lines mark $\langle N_{\mathrm{lar}} \rangle + \langle N_{\mathrm{2nd}} \rangle = 0.99 N_{\mathrm{tot}}$ and $0.999 N_{\mathrm{tot}}$. (b) Same as panel (a) but for the ejecta except for the three large remnants. The white dotted line marks $\langle N_{\mathrm{lar}} \rangle + \langle N_{\mathrm{2nd}} \rangle + \langle N_{\mathrm{3rd}} \rangle = 0.9 N_{\mathrm{tot}}$. The white dashed lines mark $\langle N_{\mathrm{lar}} \rangle + \langle N_{\mathrm{2nd}} \rangle + \langle N_{\mathrm{3rd}} \rangle = 0.99 N_{\mathrm{tot}}$ and $0.999 N_{\mathrm{tot}}$.}
  \label{fig:Ntop_vcol_N2a3_Ntar262144}
\end{figure}

We present the fraction of small fragments ejected by collisions, $[N_{\mathrm{tot}} - (\langle N_{\mathrm{lar}} \rangle + \langle N_{\mathrm{2nd}} \rangle )] / N_{\mathrm{tot}}$ [Figure \ref{fig:Ntop_vcol_N2a3_Ntar262144}-(a)] and $[N_{\mathrm{tot}} - (\langle N_{\mathrm{lar}} \rangle + \langle N_{\mathrm{2nd}} \rangle + \langle N_{\mathrm{3rd}} \rangle )] / N_{\mathrm{tot}}$ [Figure \ref{fig:Ntop_vcol_N2a3_Ntar262144}-(b)] for $N_{\mathrm{tar}} = 262144$, against $N_{\mathrm{tar}} / N_{\mathrm{pro}}$ and $v_{\mathrm{col}}$, where $N_{\mathrm{3rd}}$ is the number of monomers in the third largest remnant resulting from a collision (hereafter, called the third remnant).
The fraction of small fragments is small for unequal-mass collisions.
In particular, $\langle N_{\mathrm{lar}} \rangle + \langle N_{\mathrm{2nd}} \rangle > 0.9 N_{\mathrm{tot}}$ for collisions with the mass transfer from the target to the projectile, i.e., such unequal-mass collisions eject relatively a smaller number of small fragments even for $v_{\mathrm{col}} \gtrsim 50 ~ \mathrm{m ~ s^{-1}}$.
Target disruption producing many small fragments does not occur even though the transfer erosion dominates.
On the other hand, nearly equal-mass collisions eject relatively a larger number of small fragments even for low $v_{\mathrm{col}}$, in which collisional growth is realized, $N_{\mathrm{lar}} > N_{\mathrm{tar}}$.

In realistic PPD conditions, the collision velocity is typically $\lesssim 10^2 ~ \mathrm{m ~ s^{-1}}$.
For such a collision velocity, $\langle N_{\mathrm{lar}} \rangle + \langle N_{\mathrm{2nd}} \rangle + \langle N_{\mathrm{3rd}} \rangle \gtrsim 0.9 N_{\mathrm{tot}}$, and most of dust monomers are included in the three large remnants independently of the mass transfer from the target to the projectile.
The rough fitting formulae of the lines in Figure \ref{fig:Ntop_vcol_N2a3_Ntar262144}-(a) are
\begin{eqnarray*}
  v_{\mathrm{col}} \approx \left \{ \begin{array}{l}
    15 \times (N_{\mathrm{tar}} / N_{\mathrm{pro}})^{1.1} ~ \mathrm{m ~ s^{-1}} \\
    7.9 \times (N_{\mathrm{tar}} / N_{\mathrm{pro}})^{0.67} ~ \mathrm{m ~ s^{-1}} \\
    7.8 \times (N_{\mathrm{tar}} / N_{\mathrm{pro}})^{0.40} ~ \mathrm{m ~ s^{-1}}
  \end{array} \right . \left . \begin{array}{l}
    \mathrm{for} ~ \langle N_{\mathrm{lar}} \rangle + \langle N_{\mathrm{2nd}} \rangle = 0.9 N_{\mathrm{tot}} \mathrm{,} \\
    \mathrm{for} ~ \langle N_{\mathrm{lar}} \rangle + \langle N_{\mathrm{2nd}} \rangle = 0.99 N_{\mathrm{tot}} \mathrm{,} \\
    \mathrm{for} ~ \langle N_{\mathrm{lar}} \rangle + \langle N_{\mathrm{2nd}} \rangle = 0.999 N_{\mathrm{tot}} \mathrm{,}
  \end{array} \right .
  \label{equ:small_fragments_fitting_2nd}
\end{eqnarray*}
and similarly the fitting formulae in Figure \ref{fig:Ntop_vcol_N2a3_Ntar262144}-(b) are
\begin{eqnarray*}
  v_{\mathrm{col}} \approx \left \{ \begin{array}{l}
    70 \times (N_{\mathrm{tar}} / N_{\mathrm{pro}})^{0.51} ~ \mathrm{m ~ s^{-1}} \\
    26 \times (N_{\mathrm{tar}} / N_{\mathrm{pro}})^{0.37} ~ \mathrm{m ~ s^{-1}} \\
    21 \times (N_{\mathrm{tar}} / N_{\mathrm{pro}})^{0.24} ~ \mathrm{m ~ s^{-1}}
  \end{array} \right . \left . \begin{array}{l}
    \mathrm{for} ~ \langle N_{\mathrm{lar}} \rangle + \langle N_{\mathrm{2nd}} \rangle + \langle N_{\mathrm{3rd}} \rangle = 0.9 N_{\mathrm{tot}} \mathrm{,} \\
    \mathrm{for} ~ \langle N_{\mathrm{lar}} \rangle + \langle N_{\mathrm{2nd}} \rangle + \langle N_{\mathrm{3rd}} \rangle = 0.99 N_{\mathrm{tot}} \mathrm{,} \\
    \mathrm{for} ~ \langle N_{\mathrm{lar}} \rangle + \langle N_{\mathrm{2nd}} \rangle + \langle N_{\mathrm{3rd}} \rangle = 0.999 N_{\mathrm{tot}} \mathrm{.}
  \end{array} \right .
  \label{equ:small_fragments_fitting_3rd}
\end{eqnarray*}

\begin{figure}[t]
  \plottwo{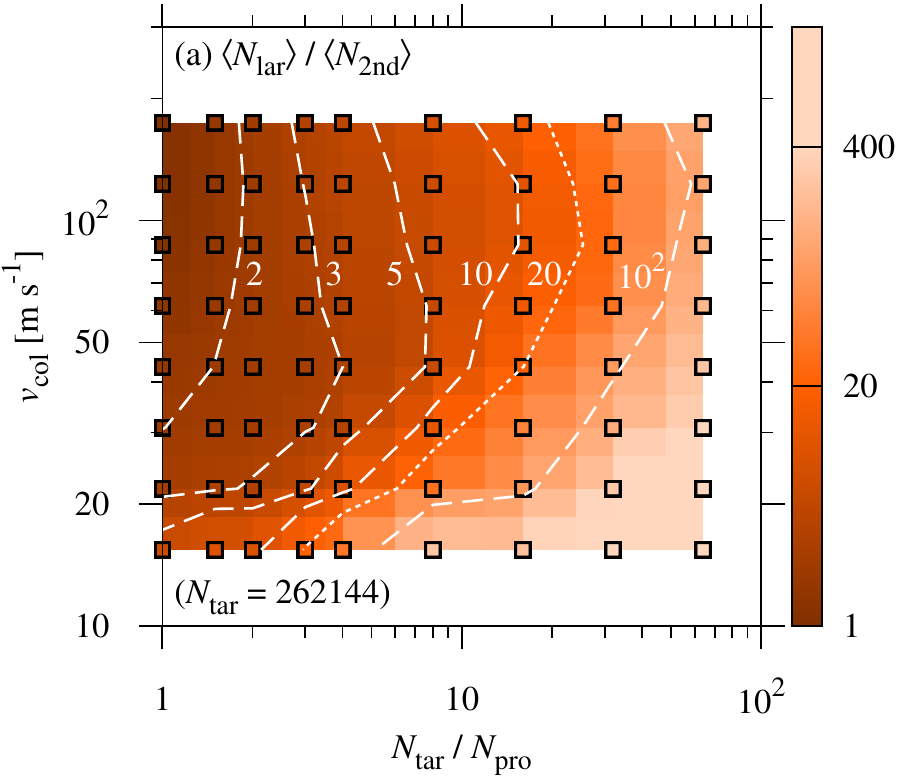}{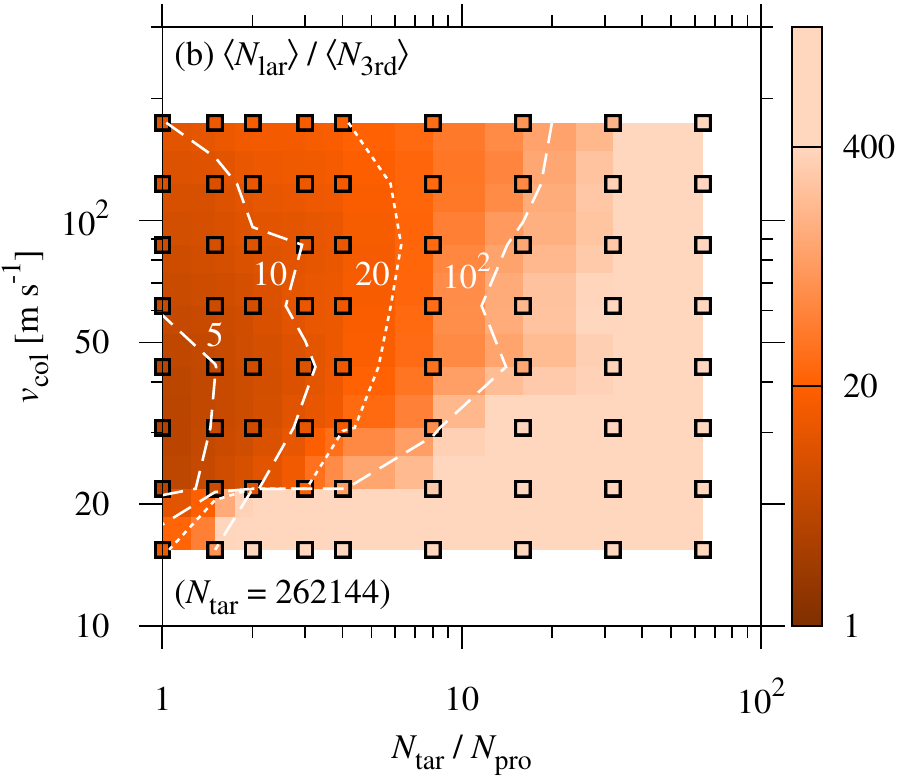}
  \caption{Ratios of largest-to-second (a) and largest-to-third (b) remnant monomer numbers on the $N_{\mathrm{tar}} / N_{\mathrm{pro}}$-$v_{\mathrm{col}}$ plane for $N_{\mathrm{tar}} = 262144$ (colors and squares). The squares are the same as Figure \ref{fig:Ntop_vcol_f1st_Ntar262144}. White lines indicate contour lines of representative ratios.}
  \label{fig:Ntop_vcol_No23_Ntar262144}
\end{figure}

As described above, our results suggest that the critical collisional fragmentation velocity for $1 < N_{\mathrm{tar}} / N_{\mathrm{pro}} \lesssim 20$ is lower than that for $N_{\mathrm{tar}} / N_{\mathrm{pro}} = 1$ owing to the mass transfer.
When we follow the growth of dust aggregates by tracking successive collisions, it is important to know mass ratios of large remnants produced by previous collisions.
Figure \ref{fig:Ntop_vcol_No23_Ntar262144} shows the relation among the numbers of monomers in the largest, second and third remnants against $N_{\mathrm{tar}} / N_{\mathrm{pro}}$ and $v_{\mathrm{col}}$ for $N_{\mathrm{tar}} = 262144$.
We focus on production of remnants with the monomer numbers to be more than $1/20$ times as large as that of the largest remnant.
The second remnant with $\langle N_{\mathrm{2nd}} \rangle \ge \langle N_{\mathrm{lar}} \rangle / 20$ is produced from collisions with $N_{\mathrm{tar}} / N_{\mathrm{pro}} \lesssim 20$.
Figure \ref{fig:Ntop_vcol_No23_Ntar262144}-(b) shows that collisions with $N_{\mathrm{tar}} / N_{\mathrm{pro}} \lesssim 6$ produce the third remnant with $\langle N_{\mathrm{3rd}} \rangle > \langle N_{\mathrm{lar}} \rangle / 20$.
In the cases that yield $\langle N_{\mathrm{3rd}} \rangle \ge \langle N_{\mathrm{lar}} \rangle / 20$, there are at least three remnants with $\langle N_{\mathrm{lar}} \rangle / 20$, including the largest remnant, whereas we have not examined the fourth largest remnant.
The minimum collision velocities to give $\langle N_{\mathrm{2nd}} \rangle \ge \langle N_{\mathrm{lar}} \rangle / 20$ and $\langle N_{\mathrm{3rd}} \rangle \ge \langle N_{\mathrm{lar}} \rangle / 20$ increase with increasing $N_{\mathrm{tar}} / N_{\mathrm{pro}}$ for $v_{\mathrm{col}} \lesssim 10^2 ~ \mathrm{m ~ s^{-1}}$
\added{
, and the rough fitting formulae of such velocities are
\begin{eqnarray}
  v_{\mathrm{col}} \approx \left \{ \begin{array}{l}
    8.6 \times (N_{\mathrm{tar}} / N_{\mathrm{pro}})^{0.55} ~ \mathrm{m ~ s^{-1}} \\
    0.26 \times (N_{\mathrm{tar}} / N_{\mathrm{pro}})^{1.9} ~ \mathrm{m ~ s^{-1}}
  \end{array} \right . \left . \begin{array}{l}
    \mathrm{for} ~ \langle N_{\mathrm{2nd}} \rangle = \langle N_{\mathrm{lar}} \rangle / 20 ~ \mathrm{and} ~ N_{\mathrm{tar}} / N_{\mathrm{pro}} \lesssim 10 \mathrm{,} \\
    \mathrm{for} ~ \langle N_{\mathrm{2nd}} \rangle = \langle N_{\mathrm{lar}} \rangle / 20 ~ \mathrm{and} ~ N_{\mathrm{tar}} / N_{\mathrm{pro}} \gtrsim 10 \mathrm{,}
  \end{array} \right .
  \label{equ:20_fitting_2nd}
\end{eqnarray}
and
\begin{eqnarray}
  v_{\mathrm{col}} \approx \left \{ \begin{array}{l}
    16 \times (N_{\mathrm{tar}} / N_{\mathrm{pro}})^{0.41} ~ \mathrm{m ~ s^{-1}} \\
    0.37 \times (N_{\mathrm{tar}} / N_{\mathrm{pro}})^{3.1} ~ \mathrm{m ~ s^{-1}}
  \end{array} \right . \left . \begin{array}{l}
    \mathrm{for} ~ \langle N_{\mathrm{3rd}} \rangle = \langle N_{\mathrm{lar}} \rangle / 20 ~ \mathrm{and} ~ N_{\mathrm{tar}} / N_{\mathrm{pro}} \lesssim 4 \mathrm{,} \\
    \mathrm{for} ~ \langle N_{\mathrm{3rd}} \rangle = \langle N_{\mathrm{lar}} \rangle / 20 ~ \mathrm{and} ~ N_{\mathrm{tar}} / N_{\mathrm{pro}} \gtrsim 4 \mathrm{.}
  \end{array} \right .
  \label{equ:20_fitting_3rd}
\end{eqnarray}
}
Okuzumi et al. (2012) showed that the growth of the target without fragmentation is mainly dominated by collisions with similar-sized projectiles ($N_{\mathrm{pro}} \gtrsim 0.1 N_{\mathrm{tar}}$) for the case with perfect sticking upon collisions.
They referred to $N$-body collision experiments by Wada et al. (2009), which showed that icy dust aggregates have high $v_{\mathrm{fra}}$, and neglected collisional fragmentation outside the snow line in PPDs.
However, our results with fragmentation suggest that, even for low $v_{\mathrm{col}}$, collisional fragmentation of icy dust aggregates occurs in collisions with $N_{\mathrm{tar}} / N_{\mathrm{pro}} > 1$.
Even though dust aggregates grow in nearly equal-mass collisions with $N_{\mathrm{tar}} / N_{\mathrm{pro}} \approx 1$, such collisions produce some large remnants that can cause collisional fragmentation in the successive collision even for low $v_{\mathrm{col}}$.

In Figure \ref{fig:Ntop_vcol_No23_Ntar262144}-(a), there is a region with $\langle N_{\mathrm{lar}} \rangle / \langle N_{\mathrm{2nd}} \rangle < N_{\mathrm{tar}} / N_{\mathrm{pro}}$.
This region roughly corresponds to the region with the mass transfer shown in Figure \ref{fig:Ntop_vcol_f2nd_Ntar262144}.
Such a collision decreases the mass ratio between the largest body and the second largest body before and after the collision.

\subsection{Summary of collisional outcomes} \label{sec:res_sum}

\begin{figure}[t]
  \plotone{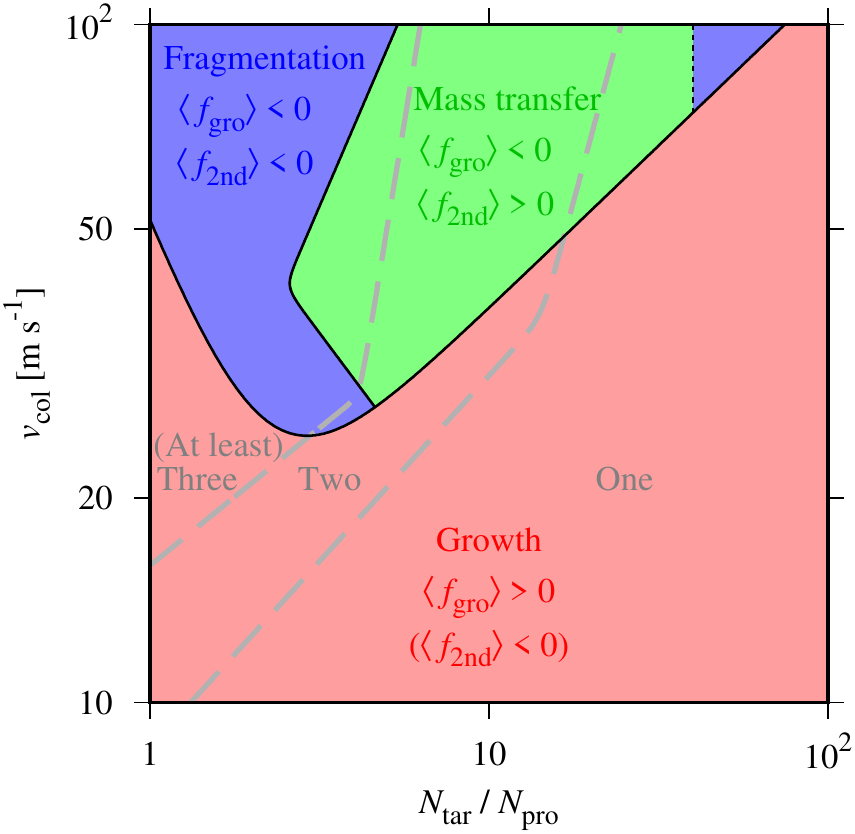}
  \caption{Summary of collision outcomes. The red, blue and green shaded regions represent the collisional growth of the target ($\langle f_{\mathrm{gro}} \rangle > 0$), the collisional fragmentation ($\langle f_{\mathrm{gro}} \rangle < 0$ and $\langle f_{\mathrm{2nd}} \rangle < 0$) and the mass transfer from the target to the projectile ($\langle f_{\mathrm{gro}} \rangle < 0$ and $\langle f_{\mathrm{2nd}} \rangle > 0$), respectively. 
\added{
The formula of the boundary between red and other regions is $v_{\mathrm{fra}} \approx (v_{\mathrm{gl}}^2 + v_{\mathrm{gh}}^2)^{1/2}$, where $v_{\mathrm{gl}} = 50 \times (N_{\mathrm{tar}} / N_{\mathrm{pro}})^{-1.3} ~ \mathrm{m ~ s^{-1}}$ and $v_{\mathrm{gh}} = 13 \times (N_{\mathrm{tar}} / N_{\mathrm{pro}})^{0.48} ~ \mathrm{m ~ s^{-1}}$ [Equations (\ref{equ:critical_collisional_growth_velocity_fit}) to (\ref{equ:critical_collisional_growth_velocity_fit_high}) in Appendix \ref{sec:app_B}]. The formulae of the left boundary between the green and blue regions are $76 \times (N_{\mathrm{tar}} / N_{\mathrm{pro}})^{-0.67} ~ \mathrm{m ~ s^{-1}}$ for $v_{\mathrm{col}} \lesssim 40 ~ \mathrm{m ~ s^{-1}}$ and $14 \times (N_{\mathrm{tar}} / N_{\mathrm{pro}})^{1.2} ~ \mathrm{m ~ s^{-1}}$ for $v_{\mathrm{col}} \gtrsim 40 ~ \mathrm{m ~ s^{-1}}$ [Equation (B4); Appendix \ref{sec:app_B}]. 
} \replaced{ The rightward boundary}{
 The right boundary, $N_{\mathrm{tar}} / N_{\mathrm{pro}} \approx 40$,
}
 between the green and blue regions is shown by the dotted line instead of a solid line because of uncertainty (Appendix \ref{sec:app_B}). The region is also classified into three areas by the number of large remnants [gray dashed lines 
\added{
given by Equations (4) and (5)
}
]. From left to right, the number of remnants decreases from three ($\langle N_{\mathrm{2nd}} \rangle$, $\langle N_{\mathrm{3rd}} \rangle > \langle N_{\mathrm{lar}} \rangle / 20$), two ($\langle N_{\mathrm{2nd}} \rangle > \langle N_{\mathrm{lar}} \rangle / 20$ and $\langle N_{\mathrm{3rd}} \rangle < \langle N_{\mathrm{lar}} \rangle / 20$), and one ($\langle N_{\mathrm{2nd}} \rangle < \langle N_{\mathrm{lar}} \rangle / 20$).}
  \label{fig:Ntop_vcol}
\end{figure}

We summarized the outcomes of collisions with various $N_{\mathrm{tar}} / N_{\mathrm{col}}$ and $v_{\mathrm{col}}$ in Figure \ref{fig:Ntop_vcol}, which are obtained from our results of Figures \ref{fig:Ntop_vcol_f1st_Ntar262144}, \ref{fig:Ntop_vcol_f2nd_Ntar262144}, and \ref{fig:Ntop_vcol_No23_Ntar262144} (see also Appendix \ref{sec:app_B}).
The obtained $v_{\mathrm{fra}}$ for unequal-mass collisions with $1 < N_{\mathrm{tar}} / N_{\mathrm{pro}} \lesssim 20$ is lower than $v_{\mathrm{fra}} \approx 50 ~ \mathrm{m ~ s^{-1}}$ with $N_{\mathrm{tar}} / N_{\mathrm{pro}} = 1$ owing to the mass transfer from the target to the projectile.
Readers clearly see the region of the mass transfer, which is drawn from Figure \ref{fig:Ntop_vcol_f1st_Ntar262144}-(b) and Figure \ref{fig:Ntop_vcol_f2nd_Ntar262144}, in $v_{\mathrm{col}} > 20 ~ \mathrm{m ~ s^{-1}}$ and $3 \lesssim N_{\mathrm{tar}} / N_{\mathrm{pro}} \lesssim 30$, whereas the upper bound of $N_{\mathrm{tar}} / N_{\mathrm{pro}}$ is still uncertain (black dotted line; Appendix \ref{sec:app_B}).
The collision with $N_{\mathrm{tar}} / N_{\mathrm{pro}} \approx$ 3-5 and $v_{\mathrm{col}} \gtrsim 30 ~ \mathrm{m ~ s^{-1}}$, which induces mass transfer, produces at least three large remnants.
For the very low collision velocity, $v_{\mathrm{col}} \lesssim 10 ~ \mathrm{m ~ s^{-1}}$, few large remnants will be ejected even for the low mass ratio.

\section{Discussion} \label{sec:dis}

\subsection{Dust growth process in protoplanetary disks} \label{sec:dis_vcol}

Utilizing Figure \ref{fig:Ntop_vcol}, we discuss the evolution of dust aggregates in the realistic condition of PPDs.
The collision velocity depends on the physical properties of PPDs in addition to the internal structure of dust aggregates.
There are various sources that cause the velocity dispersion in PPDs; the main sources are radial and azimuthal drift of dust aggregates, $\Delta v_r$ and $\Delta v_{\phi}$, and gas turbulence, $\Delta v_\mathrm{t}$, when the dust aggregates are large enough to be affected by collisional fragmentation (Krijt et al. 2015).
Then, the collision velocity is estimated to be $v_{\mathrm{col}} \approx [(\Delta v_r)^2 + (\Delta v_{\phi})^2 + (\Delta v_\mathrm{t})^2]^{1/2}$.

The Stokes number of a dust aggregate is given by $\mathrm{St} = {\Omega}_{\mathrm{K}} t_{\mathrm{stop}}$, where ${\Omega}_{\mathrm{K}}$ is the Keplerian angular velocity and $t_{\mathrm{stop}}$ is the stopping time of the dust aggregate (Birnstiel et al. 2010, 2012; Takeuchi et al. 2012; Okuzumi et al. 2012); the stopping time is in proportion to ${\rho}_{\mathrm{s}} s$ in the Epstein regime and ${\rho}_{\mathrm{s}} s^2$ in the Stokes regime, respectively, where ${\rho}_{\mathrm{s}}$ is the internal density and $s$ is the radius of a dust aggregate.
When ${\rho}_{\mathrm{s}}$ of the target and the projectile are the same, the Stokes number of the projectile, ${\mathrm{St}}_{\mathrm{pro}}$, is given by
\begin{eqnarray}
  {\mathrm{St}}_{\mathrm{pro}} = \left \{ \begin{array}{ll}
    (N_{\mathrm{tar}} / N_{\mathrm{pro}})^{-1/3} {\mathrm{St}}_{\mathrm{tar}} & ~ \textrm{for Epstein's law} \mathrm{,} \\
    (N_{\mathrm{tar}} / N_{\mathrm{pro}})^{-2/3} {\mathrm{St}}_{\mathrm{tar}} & ~ \textrm{for Stokes' law} \mathrm{,}
  \end{array} \right .
  \label{equ:Stokes}
\end{eqnarray}
respectively, where ${\mathrm{St}}_{\mathrm{tar}}$ is the Stokes number of the target.

The radial drift speed $v_r$ is given by
\begin{equation}
  v_r = \frac{2 \mathrm{St}}{1 + {\mathrm{St}}^2} \eta v_{\mathrm{K}}
  \mathrm{,}
  \label{equ:radial_drift_velocity}
\end{equation}
where $v_{\mathrm{K}}$ is the Keplerian velocity, and $\eta$ is half of the ratio between the radial gas pressure gradient force and the gravitational force of the central star (Adachi et al. 1976; Weidenschilling 1977; Nakagawa et al. 1986).
The maximum radial drift speed is $\eta v_{\mathrm{K}}$ at $\mathrm{St} = 1$.
In the MMSN model, $\eta = 1.8 \times 10^{-3} ~ (r / 1 \mathrm{au})^{1/2}$, where $r$ is the distance from the central star, and $\eta v_{\mathrm{K}} = 54 ~ \mathrm{m ~ s^{-1}}$ is independent of $r$.

The azimuthal drift speed $v_{\phi}$ is given by
\begin{equation}
  v_{\phi} = \frac{1}{1 + {\mathrm{St}}^2} \eta v_{\mathrm{K}}
  \label{equ:azimuthal_drift_velocity}
\end{equation}
(Adachi et al. 1976; Weidenschilling 1977; Nakagawa et al. 1986).
The azimuthal drift speed decreases with $\mathrm{St}$ and approaches to zero for $\mathrm{St} \gg 1$.
The maximum relative velocity is obtained for collisions between dust aggregates with ${\mathrm{St}}_{\mathrm{tar}} \sim 1$.
Then, relative velocities due to radial and azimuthal drift are
\begin{equation}
  \Delta v_r = \left | 1 - \frac{2 {\mathrm{St}}_{\mathrm{pro}}}{1 + {\mathrm{St}}_{\mathrm{pro}}^2} \right | \eta v_{\mathrm{K}}
  \mathrm{,}
  \label{equ:radial_drift_relative_velocity}
\end{equation}
and
\begin{equation}
  \Delta v_{\phi} = \left | \frac{1}{2} - \frac{1}{1 + {\mathrm{St}}_{\mathrm{pro}}^2} \right | \eta v_{\mathrm{K}}
  \mathrm{,}
  \label{equ:azimuthal_drift_relative_velocity}
\end{equation}
respectively, for ${\mathrm{St}}_{\mathrm{tar}} = 1$.

We adopt an analytic formula of the turbulence-driven relative velocity $\Delta v_\mathrm{t}$ derived by Ormel \& Cuzzi (2007).
We here consider the target with ${\mathrm{St}}_{\mathrm{tar}} = 1$, which gives the maximum relative velocity by the turbulence.
Then, the turbulence-driven relative velocity is given by
\begin{equation}
  \Delta v_\mathrm{t} = \left ( \frac{1}{2} + \frac{1}{1 + {\mathrm{St}}_{\mathrm{pro}}} \right ) ^{1/2} v_{\mathrm{g}}
  \mathrm{,}
  \label{equ:turbulence_driven_relative_velocity}
\end{equation}
where $v_{\mathrm{g}} = {\alpha}_{\mathrm{t}}^{1/2} c_{\mathrm{s}}$ is the velocity of gas with the largest turbulent eddies for turbulent strength, ${\alpha}_{\mathrm{t}}$, and isothermal sound speed, $c_{\mathrm{s}}$ (Ormel \& Cuzzi 2007).

\begin{figure}[t]
  \plottwo{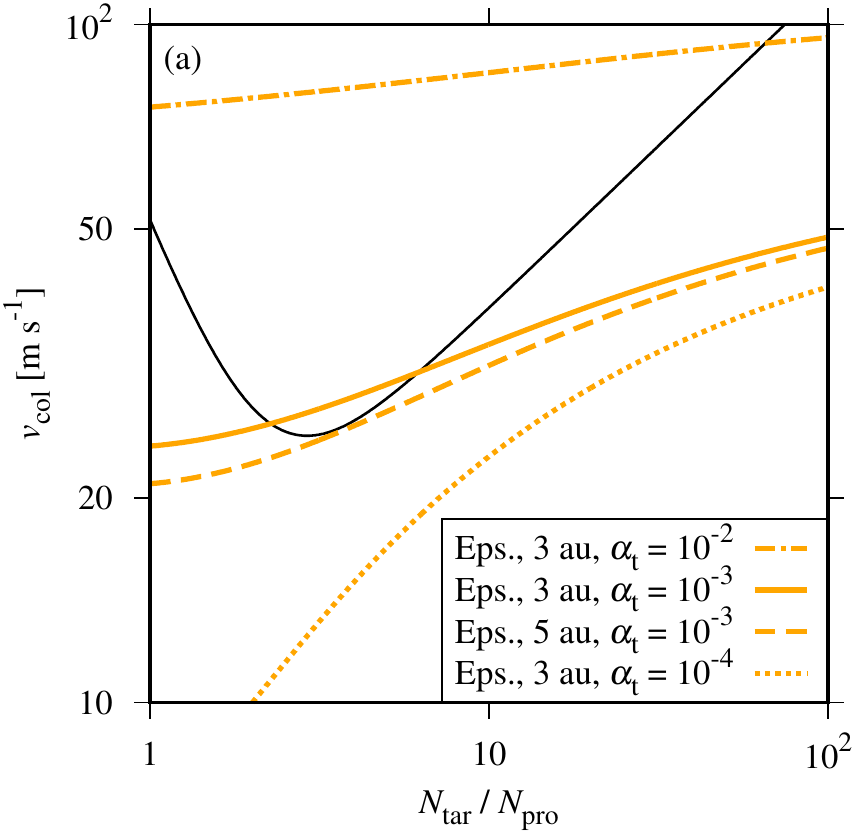}{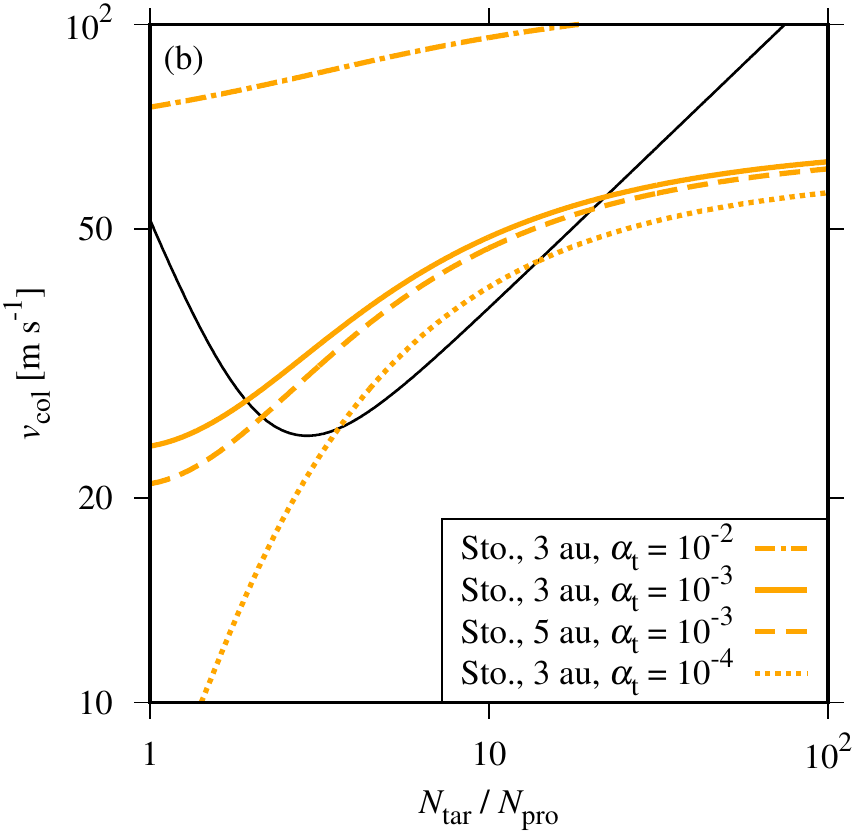}
  \caption{Collision velocity $v_{\mathrm{col}}$ against the initial mass ratio, $N_{\mathrm{tar}} / N_{\mathrm{pro}}$, for ${\mathrm{St}}_{\mathrm{tar}} = 1$ (orange solid, dashed and dotted lines). Panel (a) and (b) respectively present the results in the Epstein and Stokes regimes [Equation (\ref{equ:Stokes})]. The orange solid lines correspond to ${\alpha}_{\mathrm{t}} = 10^{-3}$ at 3 au in the MMSN ($T = 160$ K and $\eta = 3.1 \times 10^{-3}$); the orange dashed lines are for ${\alpha}_{\mathrm{t}} = 10^{-3}$ at 5 au in the MMSN ($T = 130$ K and $\eta = 4.0 \times 10^{-3}$); 
\replaced{
the orange dotted lines are for ${\alpha}_{\mathrm{t}} = 10^{-4}$ at 3 au.}{
the orange dash-dotted and dotted lines are for ${\alpha}_{\mathrm{t}} = 10^{-2}$ and $10^{-4}$ at 3 au, respectively.
}
 The black solid line is the same as the critical collisional fragmentation velocity shown in Figure \ref{fig:Ntop_vcol}.}
  \label{fig:discussion_Ntop_vcol_EaS}
\end{figure}

Figure \ref{fig:discussion_Ntop_vcol_EaS} shows the collision velocity, $v_{\mathrm{col}}$, between dust aggregates against $N_{\mathrm{tar}} / N_{\mathrm{pro}}$ for various PPD conditions.
Dust aggregates in the Stokes drag regime [Figure \ref{fig:discussion_Ntop_vcol_EaS}-(b)] have the higher collision velocity than those in the Epstein drag regime [Figure \ref{fig:discussion_Ntop_vcol_EaS}-(a)] in the entire region of $N_{\mathrm{tar}} / N_{\mathrm{pro}}$.
Figure \ref{fig:discussion_Ntop_vcol_EaS}-(b) exhibits that collision velocities (orange lines) exceed the critical collisional fragmentation velocity, $v_{\mathrm{fra}}$, (black solid line) in a wide range of the parameters; dust aggregates in the Stokes regime undergo collisional fragmentation particularly when the turbulence is strong, ${\alpha}_{\mathrm{t}} \gtrsim 10^{-3}$.
In contrast, the collision velocity in the Epstein regime is lower than $v_{\mathrm{fra}}$ [Figure \ref{fig:discussion_Ntop_vcol_EaS}-(a)], which indicates that the fragmentation barrier is not severe.
However, it is generally considered that dust aggregates in the Epstein drag regime suffer from serious inward radial drift (Okuzumi et al. 2012), whereas the radial drift barrier could be overcome if MHD disk winds are taken into account (Suzuki et al. 2016; Taki et al. 2020).

\begin{figure}[t]
  \plotone{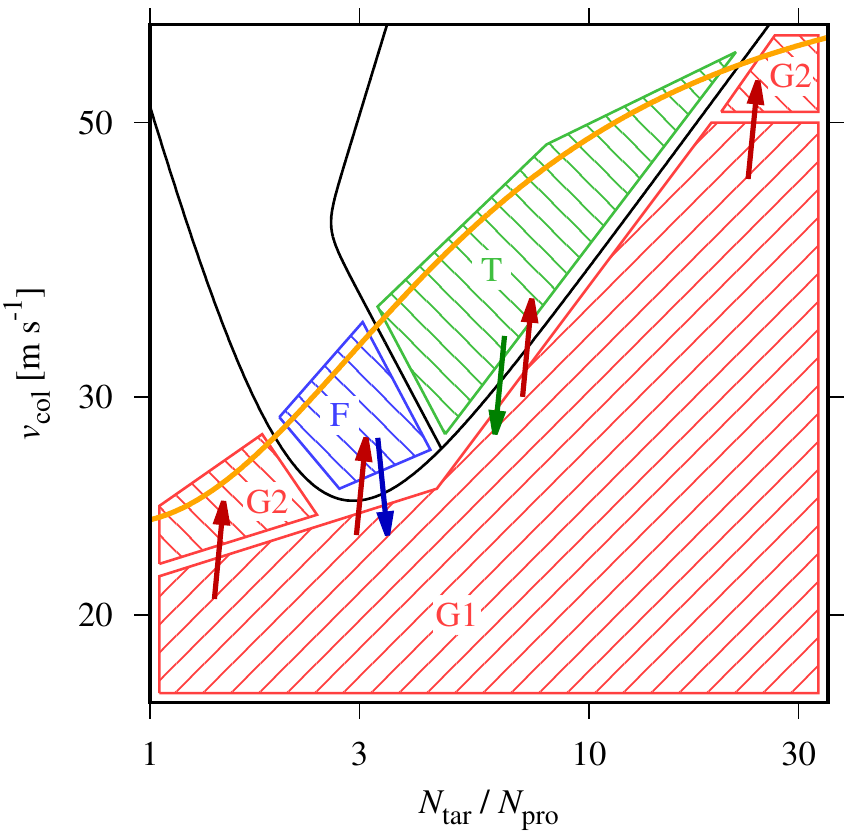}
  \caption{Schematic showing the dust evolution through collisional growth and fragmentation of dust aggregates, and mass transfer from the target to the projectile. The black lines are the same as Figure \ref{fig:Ntop_vcol}. The orange solid line is the same as Figure \ref{fig:discussion_Ntop_vcol_EaS}-(b). The red, blue and green arrows illustrate the collisional growth, the collisional fragmentation and the mass transfer, respectively. Collisional growth dominates in the regions labeled "G1" and "G2"; collisional fragmentation in the region with "F"; mass transfer in the region with "T". See text for the detail.}
  \label{fig:discussion_Ntop_vcol_N12O}
\end{figure}

Figure \ref{fig:discussion_Ntop_vcol_N12O} shows the schematic diagram of the dust evolution.
The orange solid line indicates $v_{\mathrm{col}}$ for ${\alpha}_{\mathrm{t}} = 10^{-3}$ at $r = 3$ au in the Stokes regime, which gives the maximum $v_{\mathrm{col}}$ in the six cases presented in Figure \ref{fig:discussion_Ntop_vcol_EaS}.
Collisions between small dust aggregates with $\mathrm{St} \ll 1$ have $v_{\mathrm{col}} < v_{\mathrm{fra}}$ (region G1 in Figure \ref{fig:discussion_Ntop_vcol_N12O}).
Then the target can grow through collisional sticking.
When ${\mathrm{St}}_{\mathrm{tar}}$ gets near unity due to the collisional growth of the target (red arrows in Figure \ref{fig:discussion_Ntop_vcol_N12O}), the mass ratio of colliding dust aggregates controls the properties of the next collision.
In order that the target with ${\mathrm{St}}_{\mathrm{tar}} \sim 1$ in the Stokes regime grows via collisional sticking, collisions with nearly equal mass, $N_{\mathrm{tar}} / N_{\mathrm{pro}} \lesssim 2$, or with a high mass ratio, $N_{\mathrm{tar}} / N_{\mathrm{pro}} \gtrsim 20$, are required (region G2).
For $2 \gtrsim N_{\mathrm{tar}} / N_{\mathrm{pro}} \gtrsim 20$, on the other hand, the collisional growth in the "G1" region proceeds to the region dominated by fragmentation ("F") or mass transfer ("T").
In the "F" region both target and projectile are fragmented and the mass ratio increases, $N_{\mathrm{lar}} / N_{\mathrm{2nd}} > N_{\mathrm{tar}} / N_{\mathrm{pro}}$ (blue arrow), as shown in Figure \ref{fig:Ntop_vcol_No23_Ntar262144}-(a).
In the "T" region, the target is eroded and the projectile grows to reduce the mass ratio, $N_{\mathrm{lar}} / N_{\mathrm{2nd}} < N_{\mathrm{tar}} / N_{\mathrm{pro}}$ (green arrow).
Both fragmentation and mass transfer delay collisional growth of icy dust aggregates.

Our results are partially inconsistent with the results obtained by Okuzumi et al. (2012); they show that the dust growth proceeds dominantly through collisions with $N_{\mathrm{tar}} / N_{\mathrm{pro}} \lesssim 10$ when the collisional fragmentation is neglected.
Figure \ref{fig:Ntop_vcol_f1st_Ntar262144} shows that the collisional growth efficiency is not high, $\langle f_{\mathrm{gro}} \rangle < 0.5$ in $N_{\mathrm{tar}} / N_{\mathrm{pro}} < 3$ and $v_{\mathrm{col}} > 20 ~ \mathrm{m ~ s^{-1}}$.
Our results indicate that the collisional growth does not progress in a straightforward manner but proceed slowly with suffering from partial fragmentation particularly in the inner part of PPDs.

Okuzumi et al. (2012), Kataoka et al. (2013) and Arakawa \& Nakamoto (2016) showed that the effective internal density of dust aggregate with $\mathrm{St} \sim 0.1$ is quite low when they do not undergo collisional fragmentation.
On the other hand, Okuzumi \& Tazaki (2019) suggested that a sizable fraction of the solid component should be compact dust grains in order to explain observations by ALMA (Stephens et al. 2017), which implies the importance of efficient collisional fragmentation.
From Figure \ref{fig:discussion_Ntop_vcol_EaS} we expect that small fragments are produced by unequal-mass collisions, which favorably explains the observation.

\subsection{Total monomer number of ejecta except for the largest remnant} \label{sec:dis_Neje}

\begin{figure}[t]
  \plotone{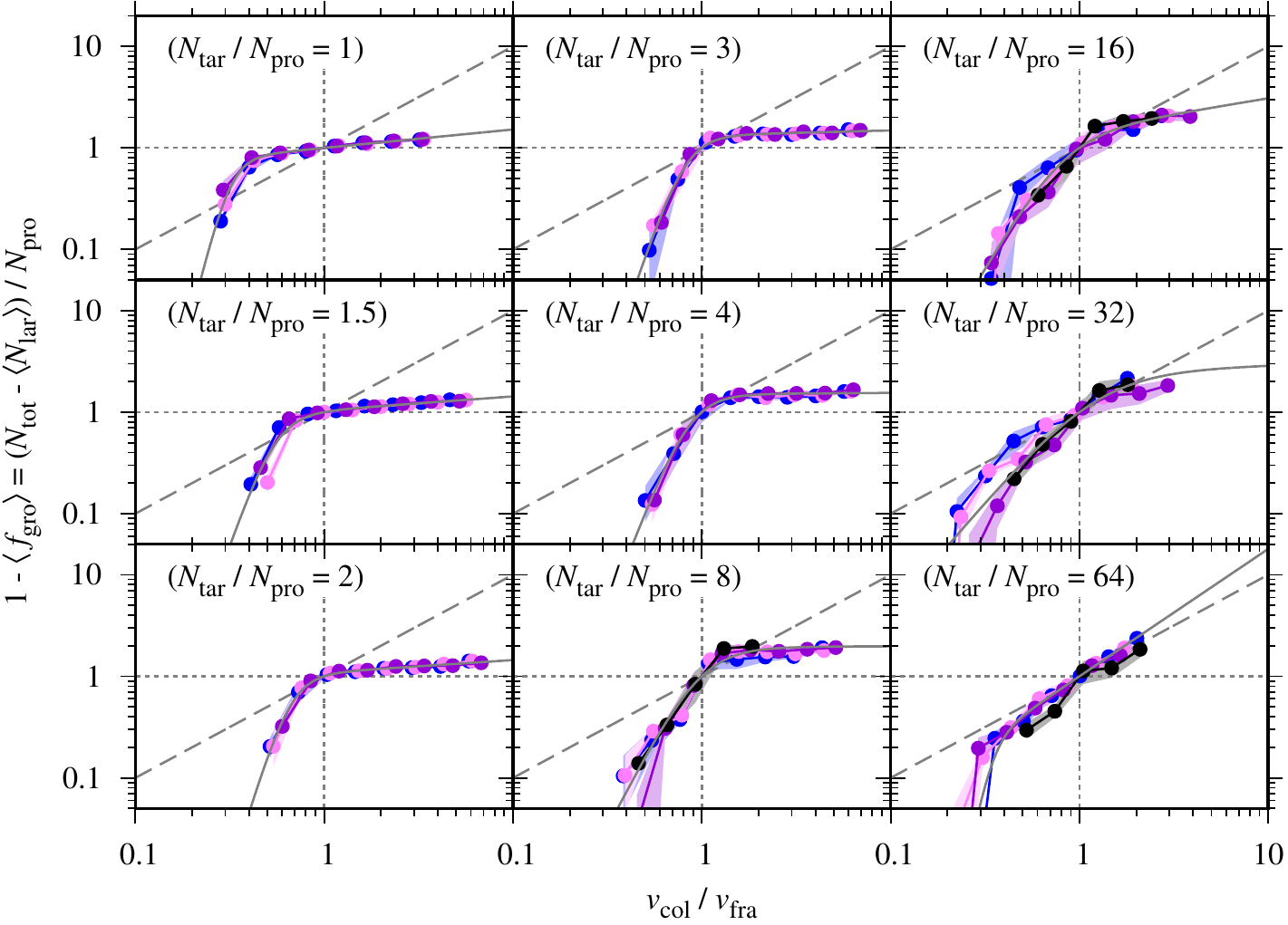}
  \caption{Total ejecta mass, except for the largest remnant, per projectile mass against the collision velocity normalized by the critical collisional fragmentation velocity (filled circles and solid lines except for gray lines). Colors represent $N_{\mathrm{tar}} =$ 65536 (blue), 131072 (magenta), 262144 (purple) and 524288 (black), respectively. Shaded regions represent the $b_{\mathrm{off}}$-weighted average standard errors of ${\bar{N}}_{\mathrm{lar}}$. Different panels correspond to results of different $N_{\mathrm{tar}} / N_{\mathrm{pro}}$. Vertical and horizontal dotted lines mark $v_{\mathrm{col}} = v_{\mathrm{fra}}$ and $\langle N_{\mathrm{lar}} \rangle = N_{\mathrm{tar}}$, respectively. Dashed lines mark $1 - \langle f_{\mathrm{gro}} \rangle = v_{\mathrm{col}} / v_{\mathrm{fra}}$. Gray solid lines mark fitting relations derived from Equation (\ref{equ:ejecta_mass_fit}).}
  \label{fig:discussion_vcoc_Meje_Ntop_multi}
\end{figure}

The reproduction of small remnants during the growth of dust particles to form planetesimals plays an important role in the planet formation though the evolution of PPDs and debris disks.
Wada et al. (2013) interpreted collisions between different-sized dust aggregates as a cratering process in dissipative media and considered a scaling relation on the cratering process (Housen \& Holsapple 2011), whereas our results show that the outcomes of collisions between dust aggregates with $3 \lesssim N_{\mathrm{tar}} / N_{\mathrm{pro}} \lesssim 30$ are the erosion with mass transfer.
In order to directly compare to these previous works, we examine the fraction of the remnants except for the largest one normalized by the projectile mass, $(N_{\mathrm{tot}} - \langle N_{\mathrm{lar}} \rangle ) / N_{\mathrm{pro}} = 1 - \langle f_{\mathrm{gro}} \rangle$.
Figure \ref{fig:discussion_vcoc_Meje_Ntop_multi} presents $1 - \langle f_{\mathrm{gro}} \rangle$ on $v_{\mathrm{col}} / v_{\mathrm{fra}}$ for different mass ratios and resolution.
Wada et al. (2013) reported that $1 - \langle f_{\mathrm{gro}} \rangle$ is roughly proportional to $v_{\mathrm{col}} / v_{\mathrm{fra}}$ for collisions with high mass ratios, $N_{\mathrm{tar}} / N_{\mathrm{pro}} =$ 16 and 64, where they assume a constant $v_{\mathrm{fra}}$ for different mass ratios.
Roughly speaking, the right three panels of Figure \ref{fig:discussion_vcoc_Meje_Ntop_multi} seem to reproduce this trend.

We fit the results of the numerical simulations in Figure \ref{fig:discussion_vcoc_Meje_Ntop_multi} by the following analytic formula,
\begin{equation}
  \left ( 1 - \langle f_{\mathrm{gro}} \rangle \right ) ^{-1} = \left [ A_1 \left ( \frac{v_{\mathrm{col}}}{v_{\mathrm{fra}}} \right ) ^{B_1} \right ] ^{-1} + \left [ A_2 \left ( \frac{v_{\mathrm{col}}}{v_{\mathrm{fra}}} \right ) ^{B_2} \right ] ^{-1}
  \mathrm{,}
  \label{equ:ejecta_mass_fit}
\end{equation}
where $A_1^{-1} + A_2^{-1} = 1$, $B_1 \le B_2$, and $v_{\mathrm{fra}}$ is derived from our simulations for each $N_{\mathrm{tar}} / N_{\mathrm{pro}}$ and each $N_{\mathrm{tar}}$.
We note that $B_1$ ($B_2$) corresponds to the power-law index for high (low) velocity collisions, $v_{\mathrm{col}} \gtrsim (\lesssim) ~ v_{\mathrm{fra}}$.

Table \ref{tab:fitting} shows $A_1$, $A_2$, $B_1$ and $B_2$ against $N_{\mathrm{tar}} / N_{\mathrm{pro}}$.
\begin{deluxetable}{ccccc}
  \tablecaption{Coefficients and power-law indices derived from Equation (\ref{equ:ejecta_mass_fit}), $A_1$, $A_2$, $B_1$ and $B_2$.}
  \label{tab:fitting}
  \tablewidth{0pt}
  \tablehead
  {
    \colhead{$N_{\mathrm{tar}} / N_{\mathrm{pro}}$} & \colhead{$A_1$} & \colhead{$A_2$} & \colhead{$B_1$} & \colhead{$B_2$}
  }
  \startdata
       1   & 1.0 & $4.3 \times 10^3$ & 0.18   & 7.5 \\
       1.5 & 1.0 & 24                & 0.14   & 5.3 \\
       2   & 1.1 & 12                & 0.12   & 6.1 \\
       3   & 1.3 &  4.0              & 0.047  & 5.5 \\
       4   & 1.5 &  3.0              & 0.012  & 4.8 \\
       8   & 1.9 &  2.1              & 0.0064 & 3.6 \\
      16   & 1.5 &  2.9              & 0.31   & 3.3 \\
      32   & 2.5 &  1.7              & 0.064  & 2.2 \\
      64   & 1.0 & $7.1 \times 10^3$ & 1.3    & 9.5 \\
  \enddata
\end{deluxetable}
For $N_{\mathrm{tar}} / N_{\mathrm{pro}} \le 32$, $B_1 \sim$ 0.01-0.1.
When the mass ratio between two colliding particles is not so high, the normalized fragmentation mass, $1 - \langle f_{\mathrm{gro}} \rangle$, does not increase so rapidly with $v_{\mathrm{col}}$ even for high velocity collisions because the projectile mass, which is in the denominator of $1 - \langle f_{\mathrm{gro}} \rangle$, is large.
Therefore, these cases do not reproduce the tendency, $1 - \langle f_{\mathrm{gro}} \rangle = v_{\mathrm{col}} / v_{\mathrm{fra}}$, obtained for collisions with high mass ratio (Wada et al. 2013).
On the other hand, for $N_{\mathrm{tar}} / N_{\mathrm{pro}} = 64$, $B_1 \approx 1$ i.e., the collision outcomes for $v_{\mathrm{col}} \gtrsim v_{\mathrm{fra}}$ roughly reproduce the proportional relation.
For all mass ratios, $B_2 > 1$, which indicates that the total ejecta mass rapidly increases with $v_{\mathrm{col}}$ in the regime of low velocity collisions, $v_{\mathrm{col}} < v_{\mathrm{fra}}$.

\section{Summary} \label{sec:sum}

The fragmentation barrier is one of the severe obstacles against the formation of planets.
The collisional fragmentation of dust aggregates hinders or slows down the growth of solid particles via collisional sticking.
When the dust growth is delayed, solid objects with $\mathrm{St} \sim 1$ suffer from inward radial drift for longer time, which may prevent them from further growing to planetesimals.
In order to clarify the growth of solid particles in PPDs, it is essential to understand the basic physical processes of collisional fragmentation, which can also be utilized in global models for the planet formation.
We examined collision outcomes between dust aggregates in the wide range of the mass ratios before collisions by carrying out $N$-body simulations of dust monomers.

For collisions between dust aggregates composed of submicron-sized icy dust monomer, the critical collisional fragmentation velocity for a pre-collisional mass ratio of $1 < N_{\mathrm{tar}} / N_{\mathrm{pro}} \lesssim 20$ is lower than that for $N_{\mathrm{tar}} / N_{\mathrm{pro}} = 1$.
The minimum critical collisional fragmentation velocity $\approx 25 ~ \mathrm{m ~ s^{-1}}$ is obtained at $N_{\mathrm{tar}} / N_{\mathrm{pro}} \approx 3$, and is about half of the critical value for equal-mass collisions.
The low critical collisional fragmentation velocity for $N_{\mathrm{tar}} / N_{\mathrm{pro}} \neq 1$ originates from offset collisions that cause the erosion of the target with the mass transfer to the projectile.
This low critical collisional fragmentation velocity may delay the collisional growth of icy dust aggregates in the Stokes drag regime, because it is considered that the dust growth is mainly dominated by collisions between dust aggregates with $N_{\mathrm{tar}} / N_{\mathrm{pro}} \lesssim 10$ if perfect sticking is assumed upon collisions (Okuzumi et al. 2012).

Applying these results to realistic PPDs, we can obtain global pictures on the growth of dust grains.
When dust aggregates are sufficiently small with $\mathrm{St} \ll 1$, they hardly suffer from fragmentation but grow via sticking.
When they grow to $\mathrm{St} \lesssim 1$, the collisional fragmentation becomes severe particularly for unequal-mass collisions with $2 \lesssim N_{\mathrm{tar}} / N_{\mathrm{pro}} \lesssim 20$; there are two possible pathways, nearly equal-mass collisions with $N_{\mathrm{tar}} / N_{\mathrm{pro}} \lesssim 2$ and collisions with a high mass ratio with $N_{\mathrm{tar}} / N_{\mathrm{pro}} \gtrsim 20$, for the growth of solid particles beyond $\mathrm{St} \gtrsim 1$.
However, nearly equal-mass collisions possibly eject multiple moderate-size remnants even for low-collision velocity, $v_{\mathrm{col}} \gtrsim 10 ~ \mathrm{m ~ s^{-1}}$, which may cause fragmentation at subsequent collisions.

In the regime of low-velocity collisions, $v_{\mathrm{col}} < v_{\mathrm{fra}}$, the total ejecta mass except for the largest remnant rapidly increases with $v_{\mathrm{col}}$.
For the regime of high-velocity collisions, on the other hand, the total ejecta mass does not increase so much with $v_{\mathrm{col}}$ for collisions between dust aggregates with low-mass ratios.

Although in this paper we handled dust aggregates that are composed of BPCA clusters, we have to cover BCCA clusters in more elaborated works because the internal structure and the mass distribution of remnants after collisions also affect the subsequent collision outcomes.
We also adopted equal-sized monomers of water ice in this paper.
The collision outcomes of dust aggregates composed of fragile silicate or $\mathrm{CO}_2$ icy dust monomers are also important to understand the dust growth in the broad regions of PPDs.
We will address those topics in future works.

\acknowledgments

We thank an anonymous referee for helpful comments.
This work was supported by Grants-in-Aid for Scientific Research from the MEXT of Japan, 17H01103, 17H01105, 17K05632, 18H05436, 18H05438, 19K03941 and 20H04612.
Our numerical computations were in part carried out on the general-purpose PC cluster at Center for Computational Astrophysics, National Astronomical Observatory of Japan.

\appendix

\section{List of sampling points in numerical input parameters} \label{sec:app_A}

\added{
As mentioned in section \ref{sec:mod}, the input parameters of our simulations are $N_{\mathrm{tar}}$, $N_{\mathrm{pro}}$, $v_{\mathrm{col}}$ and $b_{\mathrm{off}}$.
Table \ref{tab:list} lists the combination of $N_{\mathrm{tar}}$, $N_{\mathrm{pro}}$ and $v_{\mathrm{col}}$ in our simulations.
We tested eight different collision velocities except for the cases with $N_{\mathrm{tar}} = 524288$; these highest-resolution cases are performed for five different $v_{\mathrm{col}}$.
We performed simulations for ten different impact parameters from near head-on $b_{\mathrm{off}} = 0.05 b_{\mathrm{max}}$ to extreme offset $b_{\mathrm{off}} = 0.95 b_{\mathrm{max}}$ with a constant interval of $0.1 b_{\mathrm{max}}$.
The number of different sets of $(N_{\mathrm{tar}}, N_{\mathrm{pro}}, v_{\mathrm{col}}, b_{\mathrm{off}})$ is 8320, and we performed four runs for each set to give a total of 33280 runs.
}

\begin{deluxetable}{ccc}
  \tablecaption{List of sampling points in $N_{\mathrm{tar}}$, $N_{\mathrm{pro}}$ and $v_{\mathrm{col}}$.}
  \label{tab:list}
  \tablehead
  {
    \colhead{$N_{\mathrm{tar}}$} & \colhead{$N_{\mathrm{pro}}$} & \colhead{$v_{\mathrm{col}}$ [$\mathrm{m ~ s^{-1}}$]}
  }
  \startdata
    {   128} &    128                                  & 15, 22, 31, 44, 62, 87, 123, 174 \\
    \hline
    {   256} &    256,    170,    128                  & 15, 22, 31, 44, 62, 87, 123, 174 \\
    \hline
    {   512} &    512,    341,    256,    170,    128  & 15, 22, 31, 44, 62, 87, 123, 174 \\
    \hline
    {  1024} &   1024,    682,    512,    341,    256, & 15, 22, 31, 44, 62, 87, 123, 174 \\
    {      } &    128                                  & { } \\
    \hline
    {  2048} &   2048,   1365,   1024,    682,    512, & 15, 22, 31, 44, 62, 87, 123, 174 \\
    {      } &    256,    128                          & { } \\
    \hline
    {  4096} &   4096,   2730,   2048,   1365,   1024, & 15, 22, 31, 44, 62, 87, 123, 174 \\
    {      } &    512,    256,    128                  & { } \\
    \hline
    {  8192} &   8192,   5461,   4096,   2730,   2048, & 15, 22, 31, 44, 62, 87, 123, 174 \\
    {      } &   1024,    512,    256,    128          & { } \\
    \hline
    { 16384} &  16384,  10922,   8192,   5461,   4096, & 15, 22, 31, 44, 62, 87, 123, 174 \\
    {      } &   2048,   1024,    512,    256,    128  & { } \\
    \hline
    { 32768} &  32768,  21845,  16384,  10922,   8192, & 15, 22, 31, 44, 62, 87, 123, 174 \\
    {      } &   4096,   2048,   1024,    512,    256, & { } \\
    {      } &    128                                  & { } \\
    \hline
    { 65536} &  65536,  43690,  32768,  21845,  16384, & 15, 22, 31, 44, 62, 87, 123, 174 \\
    {      } &   8192,   4096,   2048,   1024,    512, & { } \\
    {      } &    256,    128                          & { } \\
    \hline
    {131072} & 131072,  87381,  65536,  43690,  32768, & 15, 22, 31, 44, 62, 87, 123, 174 \\
    {      } &  16384,   8192,   4096,   2048,   1024, & { } \\
    {      } &    512,    256,    128                  & { }\\
    \hline
    {262144} & 262144, 174762, 131072,  87381,  65536, & 15, 22, 31, 44, 62, 87, 123, 174 \\
    {      } &  32768,  16384,   8192,   4096,   2048, & { } \\
    {      } &   1024,    512,    256,    128          & { } \\
    \hline
    {524288} &  65536                                  & 15, 22, 31, 44, 62 \\
    \hline
    {524288} &  32768,  16384                          &         31, 44, 62, 87, 123 \\
    \hline
    {524288} &   8192,   4096,   2048,   1024,    512  &             44, 62, 87, 123, 174 \\
  \enddata
\end{deluxetable}

\section{Dependence of growth efficiencies on colliding dust aggregates} \label{sec:app_B}

\begin{figure}[p]
  \figurenum{11a}
  \plotone{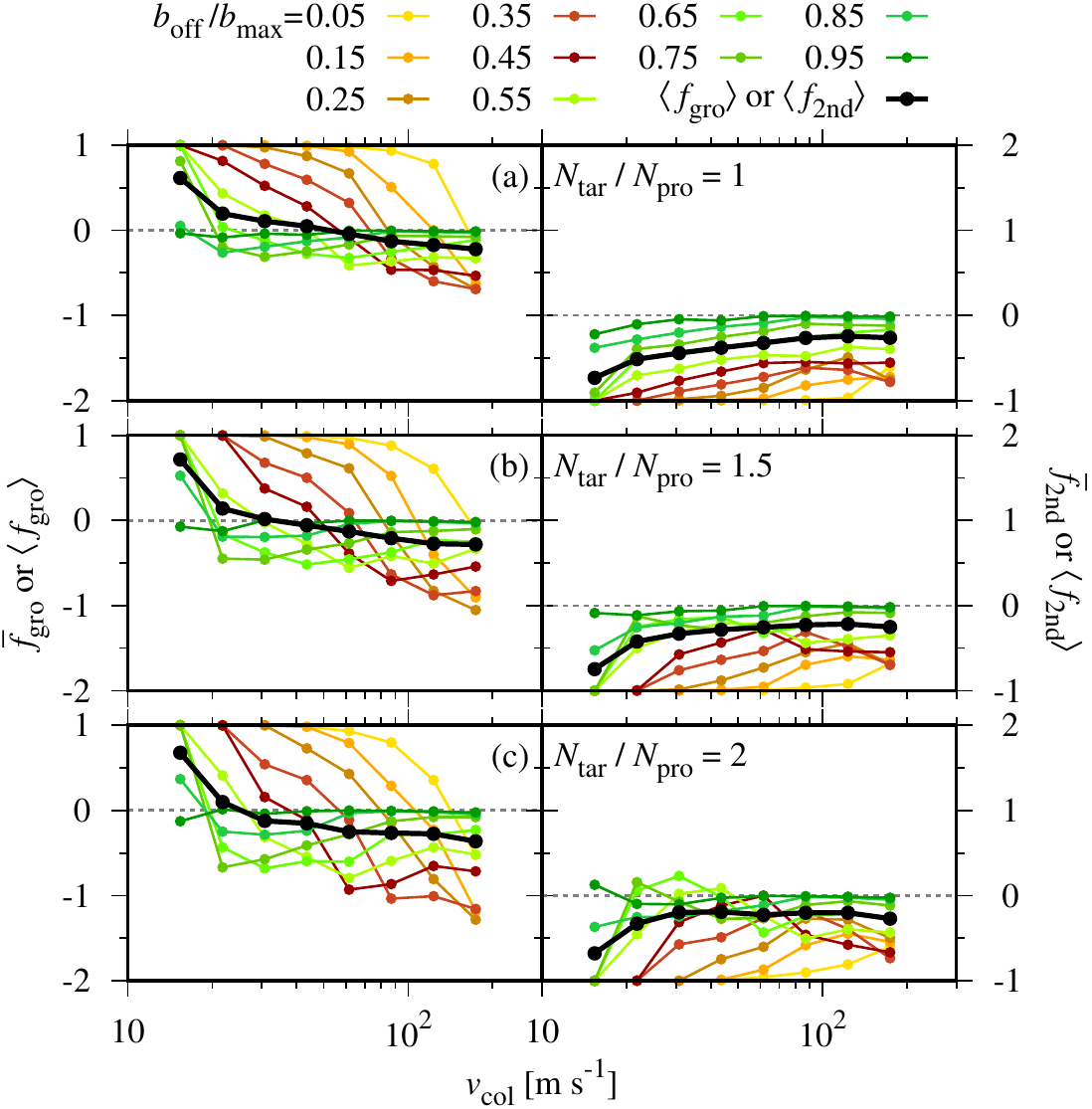}
  \caption{Average of growth efficiency of the four runs, ${\bar{f}}_{\mathrm{gro}}$ (left) or ${\bar{f}}_{\mathrm{2nd}}$ (right), against the collision velocity $v_{\mathrm{col}}$ for $N_{\mathrm{tar}} = 262144$ (filled circles and solid lines except for black ones). Colors represent the impact parameter normalized by the maximum value of the impact parameter, $b_{\mathrm{off}} / b_{\mathrm{max}}$, except for black and gray lines. The black filled circles and solid lines mark $\langle f_{\mathrm{gro}} \rangle$ (left) or $\langle f_{\mathrm{2nd}} \rangle$ (right). Gray dotted lines marks ${\bar{f}}_{\mathrm{gro}} = 0$ (left) or ${\bar{f}}_{\mathrm{2nd}} = 0$ (right). Panels are labeled by $N_{\mathrm{tar}} / N_{\mathrm{pro}} =$ 1 (a), 1.5 (b) and 2 (c), respectively.}
  \label{fig:appendix_vcol_f1f2_boff_Ntar262144_Ntop_multi_1_2}
\end{figure}

\begin{figure}[p]
  \figurenum{11b}
  \plotone{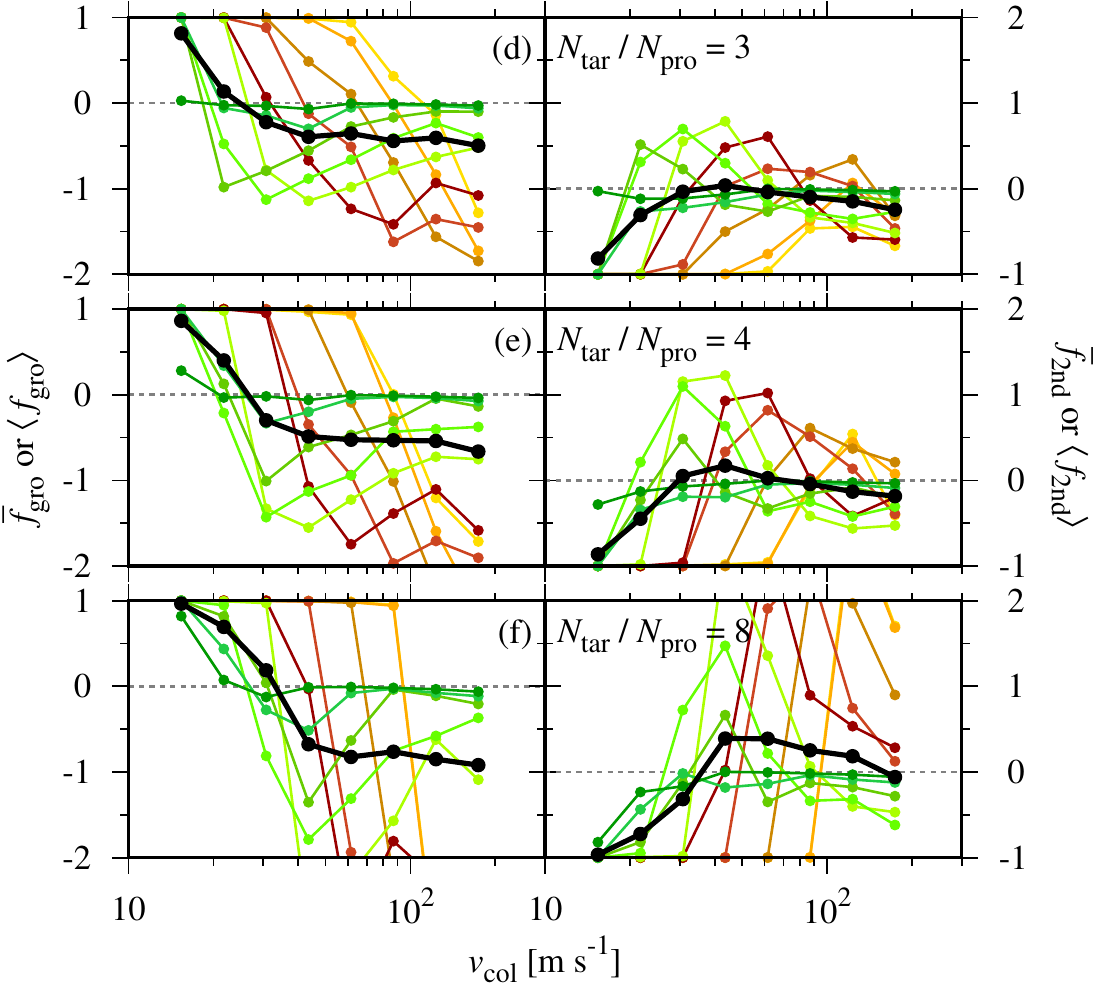}
  \caption{Same as Figure \ref{fig:appendix_vcol_f1f2_boff_Ntar262144_Ntop_multi_1_2} but for $N_{\mathrm{tar}} / N_{\mathrm{pro}} =$ 3 (d), 4 (e) and 8 (f), respectively.}
  \label{fig:appendix_vcol_f1f2_boff_Ntar262144_Ntop_multi_3_8}
\end{figure}

\begin{figure}[p]
  \figurenum{11c}
  \plotone{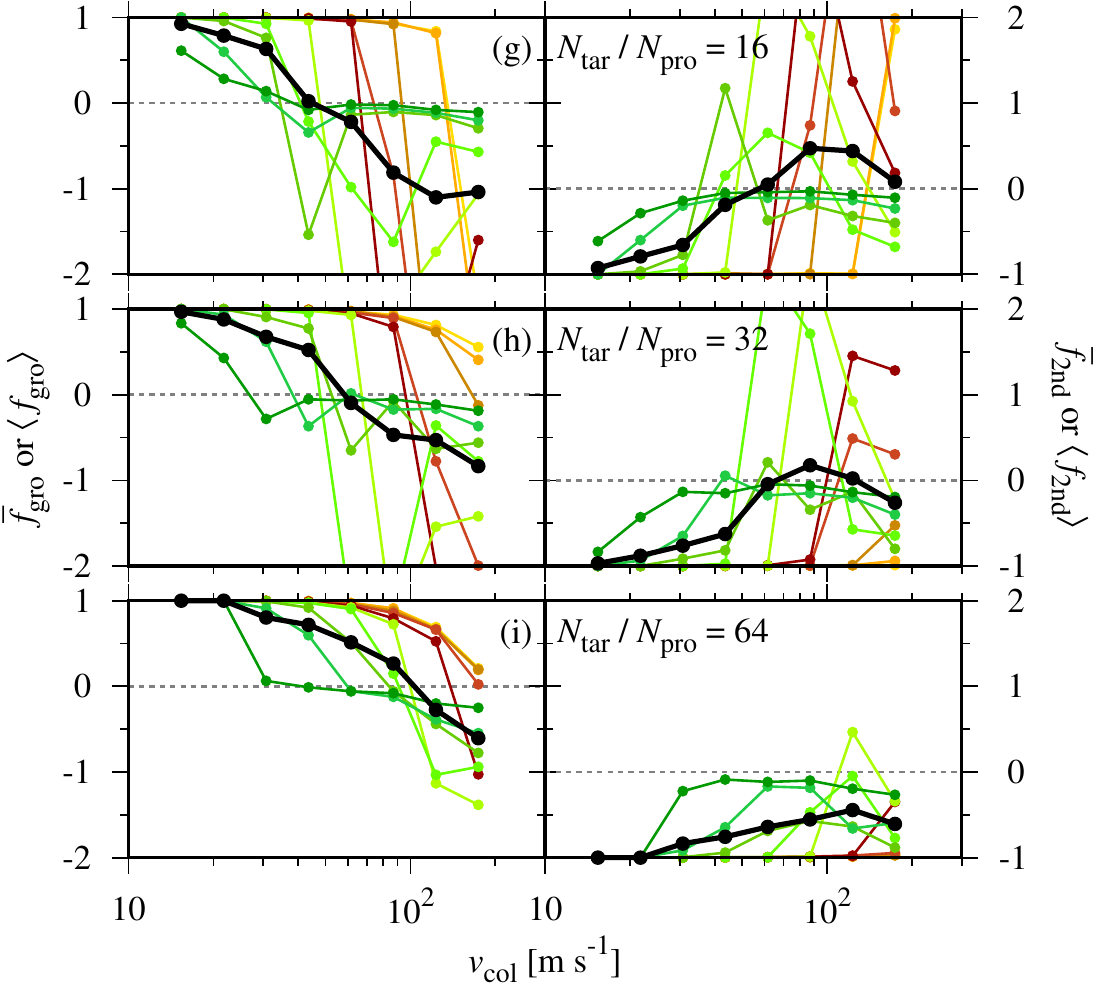}
  \caption{Same as Figure \ref{fig:appendix_vcol_f1f2_boff_Ntar262144_Ntop_multi_1_2} but for $N_{\mathrm{tar}} / N_{\mathrm{pro}} =$ 16 (g), 32 (h) and 64 (i), respectively.}
  \label{fig:appendix_vcol_f1f2_boff_Ntar262144_Ntop_multi_16_64}
\end{figure}

Figures \ref{fig:appendix_vcol_f1f2_boff_Ntar262144_Ntop_multi_1_2} to \ref{fig:appendix_vcol_f1f2_boff_Ntar262144_Ntop_multi_16_64} show ${\bar{f}}_{\mathrm{gro}}$ and ${\bar{f}}_{\mathrm{2nd}}$ used to calculate $\langle f_{\mathrm{gro}} \rangle$ and $\langle f_{\mathrm{2nd}} \rangle$ shown in Figure \ref{fig:Ntop_vcol_f1st_Ntar262144}-(a) and Figure \ref{fig:Ntop_vcol_f2nd_Ntar262144}, against $v_{\mathrm{col}}$, $b_{\mathrm{off}} / b_{\mathrm{max}}$ and $N_{\mathrm{tar}} / N_{\mathrm{pro}}$ for $N_{\mathrm{tar}} = 262144$.
These figures show the difference between head-on collisions and offset collisions.

\begin{figure}[p]
  \figurenum{12a}
  \plotone{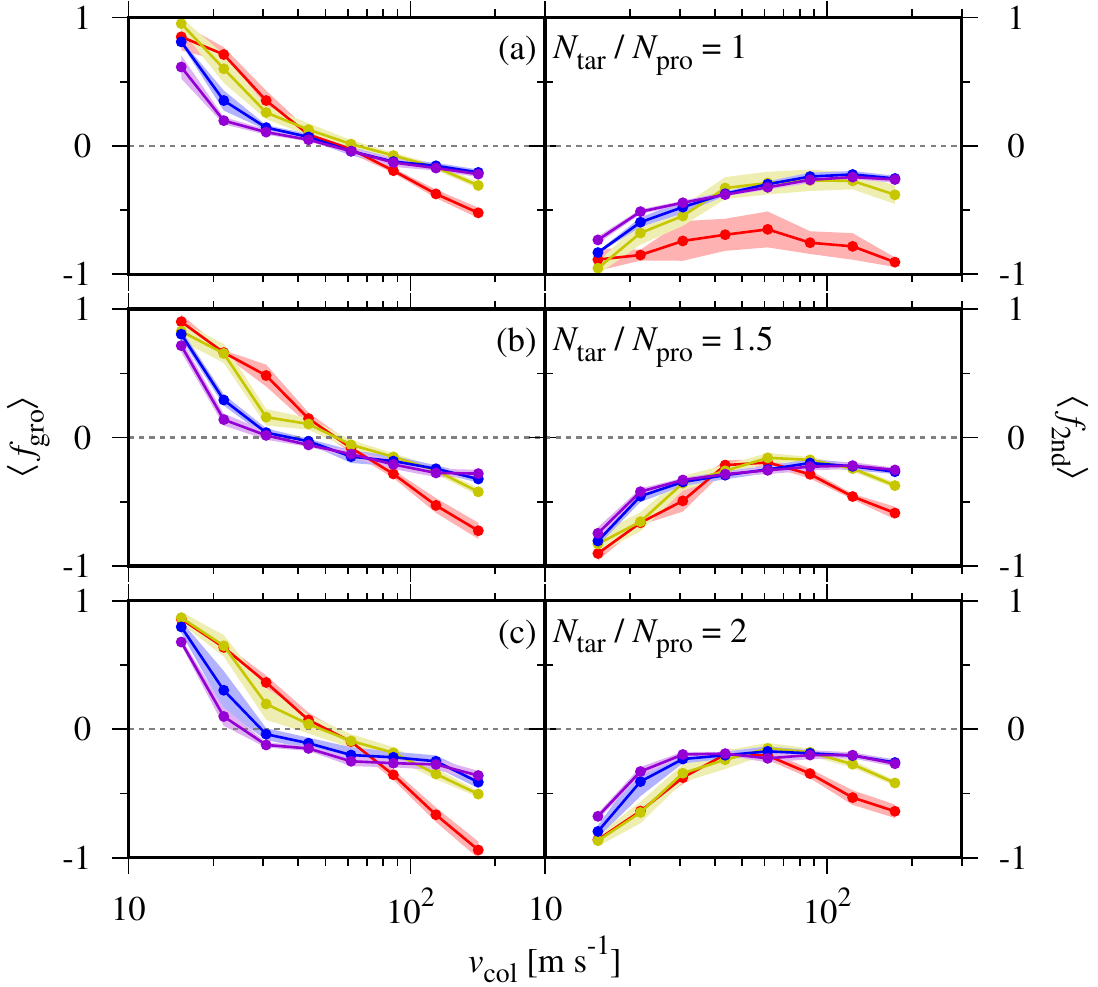}
  \caption{Growth efficiency of the $b_{\mathrm{off}}$-weighted average monomer number of the largest remnant and the second remnant, $\langle f_{\mathrm{gro}} \rangle$ (left) and $\langle f_{\mathrm{2nd}} \rangle$ (right), against $v_{\mathrm{col}}$ for $N_{\mathrm{pro}} \ge 100$ (filled circles and solid lines). Shaded regions represent the $b_{\mathrm{off}}$-weighted average standard errors of ${\bar{N}}_{\mathrm{lar}}$ (left) or ${\bar{N}}_{\mathrm{2nd}}$ (right). Colors represent $N_{\mathrm{tar}} =$ 1024 (red), 8192 (yellow), 65536 (blue) and 262144 (purple), respectively. Panels are labeled by $N_{\mathrm{tar}} / N_{\mathrm{pro}} =$ 1 (a), 1.5 (b) and 2 (c), respectively.}
  \label{fig:appendix_vcol_f1f2_Ntar_Ntop_multi_1_2}
\end{figure}

\begin{figure}[p]
  \figurenum{12b}
  \plotone{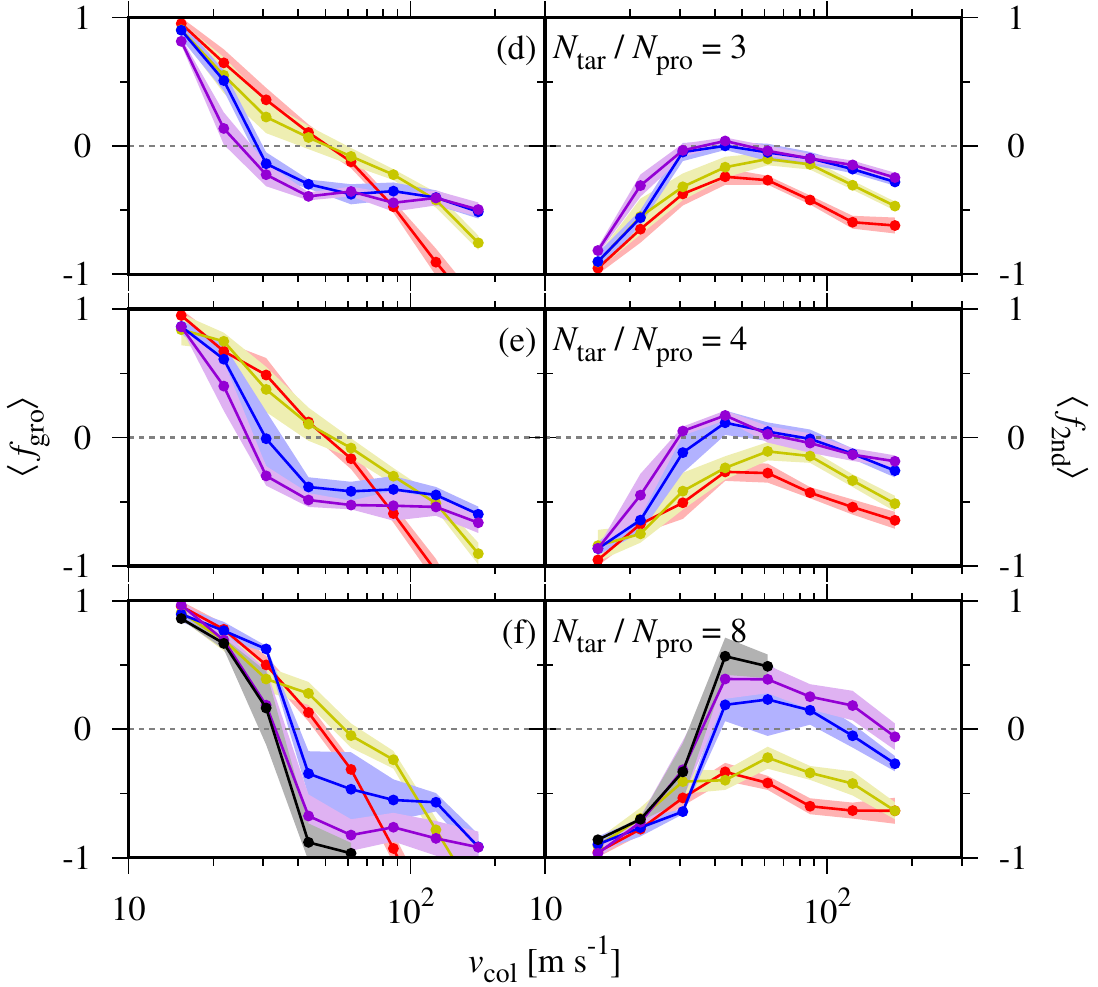}
  \caption{Same as Figure \ref{fig:appendix_vcol_f1f2_Ntar_Ntop_multi_1_2} but for $N_{\mathrm{tar}} / N_{\mathrm{pro}} =$ 3 (d), 4 (e) and 8 (f), respectively. The black filled circles, solid lines and shaded regions represent results for $N_{\mathrm{tar}} = 524288$.}
  \label{fig:appendix_vcol_f1f2_Ntar_Ntop_multi_3_8}
\end{figure}

\begin{figure}[p]
  \figurenum{12c}
  \plotone{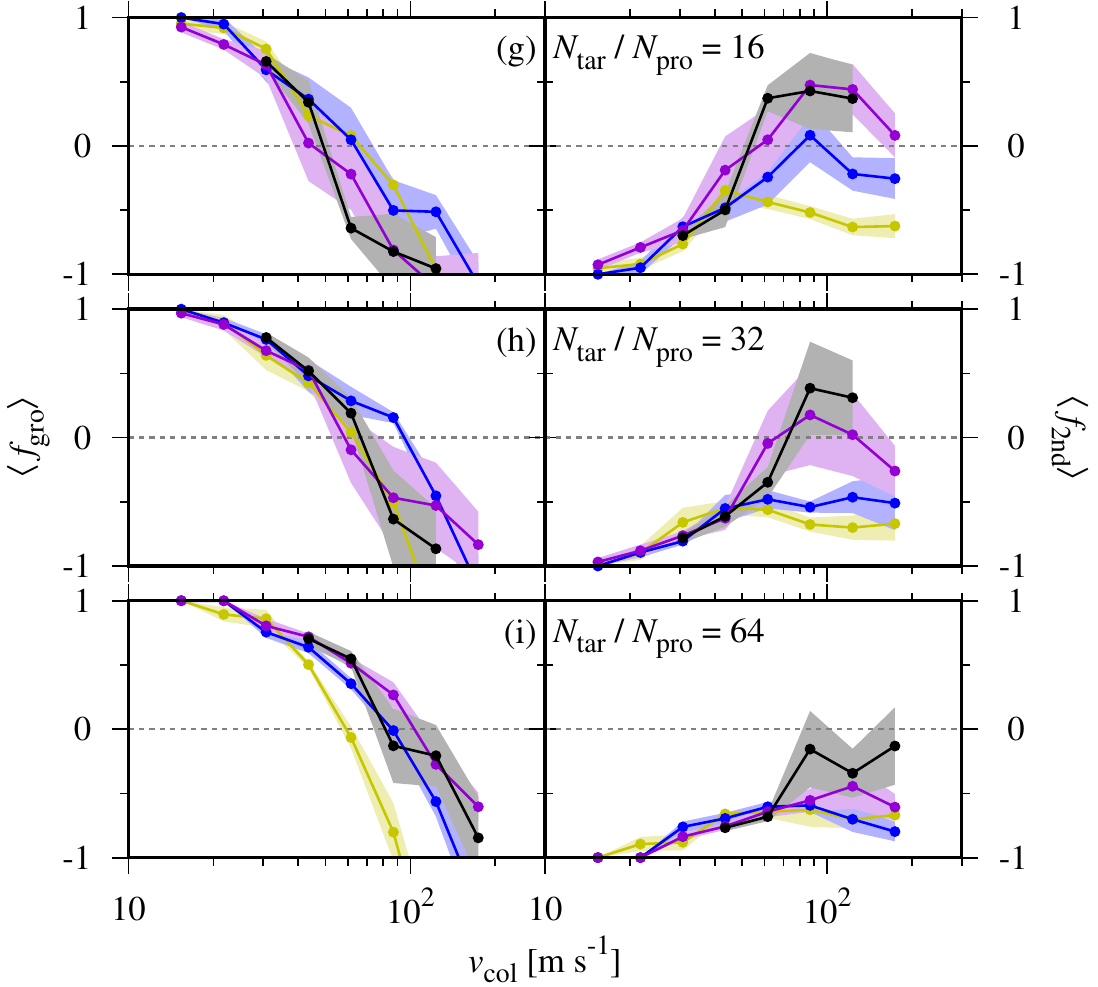}
  \caption{Same as Figure \ref{fig:appendix_vcol_f1f2_Ntar_Ntop_multi_3_8} but for $N_{\mathrm{tar}} / N_{\mathrm{pro}} =$ 16 (g), 32 (h) and 64 (i), respectively.}
  \label{fig:appendix_vcol_f1f2_Ntar_Ntop_multi_16_64}
\end{figure}

Figures \ref{fig:appendix_vcol_f1f2_Ntar_Ntop_multi_1_2} to \ref{fig:appendix_vcol_f1f2_Ntar_Ntop_multi_16_64} show $\langle f_{\mathrm{gro}} \rangle$ and $\langle f_{\mathrm{2nd}} \rangle$ against $v_{\mathrm{col}}$, $N_{\mathrm{tar}}$ and $N_{\mathrm{tar}} / N_{\mathrm{pro}}$ for $N_{\mathrm{pro}} \ge 100$.
For $N_{\mathrm{tar}} = 524288$ and $N_{\mathrm{tar}} / N_{\mathrm{pro}} = 64$ [Figure \ref{fig:appendix_vcol_f1f2_Ntar_Ntop_multi_16_64}-(i)], when taking into account the standard errors, $\langle f_{\mathrm{2nd}} \rangle$ might well be positive at $v_{\mathrm{col}} =$ 87 and $1.7 \times 10^2 ~ \mathrm{m ~ s^{-1}}$ but is still negative at $1.2 \times 10^2 ~ \mathrm{m ~ s^{-1}}$.
We suppose that $\langle f_{\mathrm{2nd}} \rangle < 0$ at $v_{\mathrm{col}} \le 10^2 ~ \mathrm{m ~ s^{-1}}$ for $N_{\mathrm{tar}} / N_{\mathrm{pro}} = 64$.

\begin{figure}[p]
  \figurenum{13}
  \plottwo{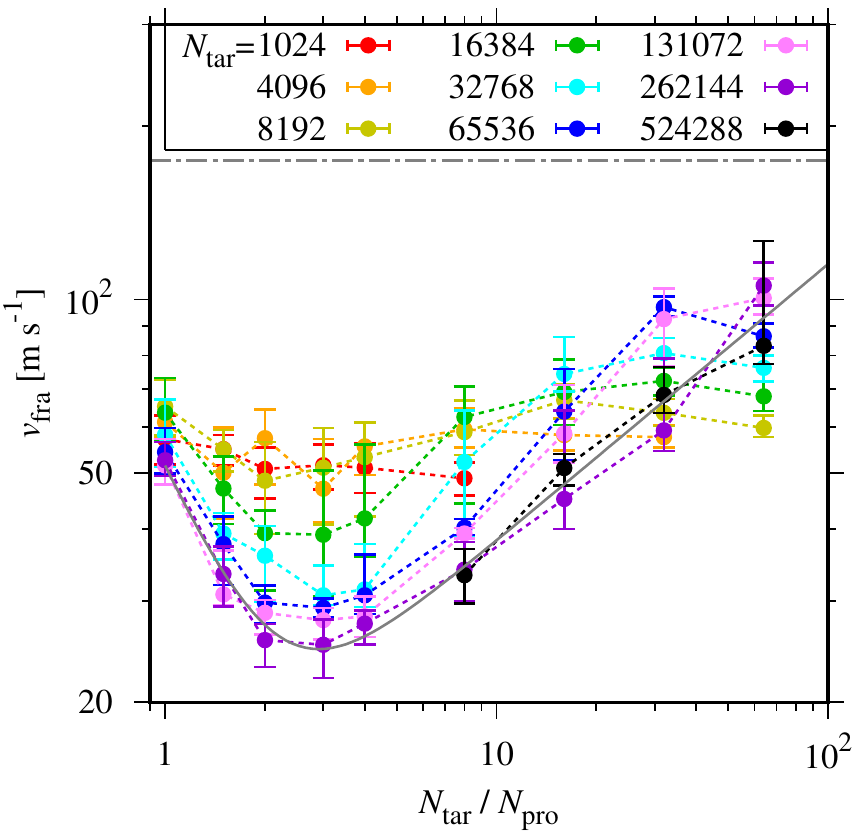}{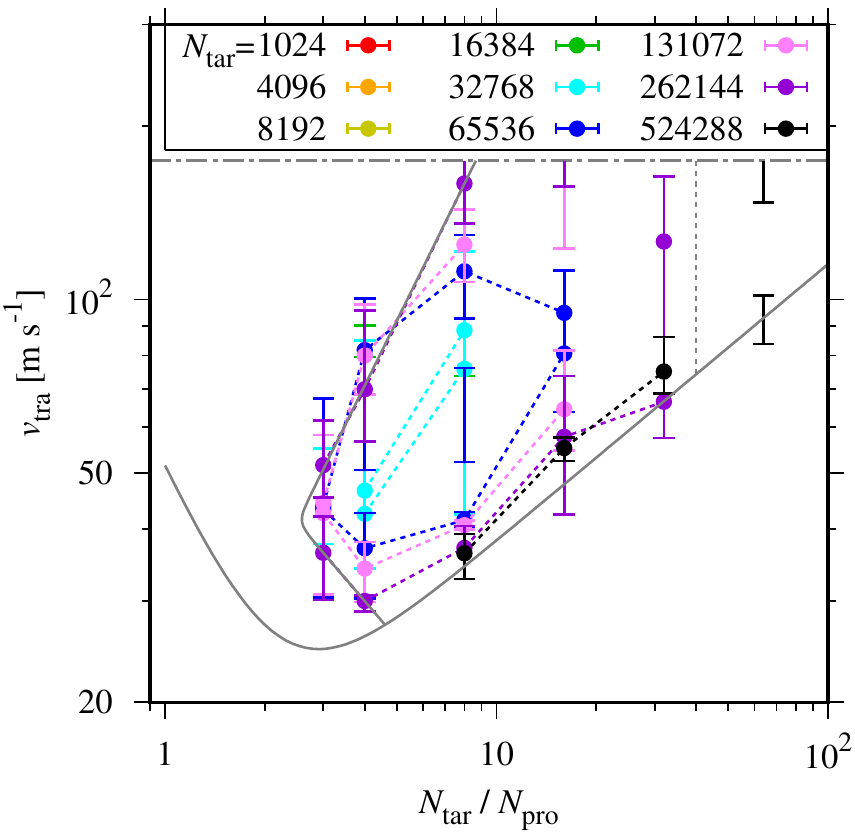}
  \caption{Critical collisional fragmentation velocity $v_{\mathrm{fra}}$ (left) and critical collisional transfer velocity $v_{\mathrm{tra}}$ (right) against the mass ratio before the collision, $N_{\mathrm{tar}} / N_{\mathrm{pro}}$ (filled circles and lines except for gray lines). Colors represent the monomer number of the target, $N_{\mathrm{tar}}$. The error bars represent the width of the critical velocity where the shaded regions and the dotted line cross in Figures \ref{fig:appendix_vcol_f1f2_Ntar_Ntop_multi_1_2} to \ref{fig:appendix_vcol_f1f2_Ntar_Ntop_multi_16_64}. The gray dash-dotted lines mark the upper end of the range of $v_{\mathrm{col}}$ in our simulations. Other gray lines are the same as the black lines in Figure \ref{fig:Ntop_vcol}.}
  \label{fig:appendix_Ntop_vcri_Ntar}
\end{figure}

Figure \ref{fig:appendix_Ntop_vcri_Ntar} shows the critical collisional fragmentation velocity $v_{\mathrm{fra}}$ (left panel) and the collision velocity with $\langle N_{\mathrm{2nd}} \rangle = N_{\mathrm{pro}}$, hereafter called the critical collisional transfer velocity, $v_{\mathrm{tra}}$ (right panel), against $N_{\mathrm{tar}} / N_{\mathrm{pro}}$ and $N_{\mathrm{tar}}$ for $N_{\mathrm{pro}} \ge 100$.
\replaced{The rough fitting formula of these $v_{\mathrm{fra}}$ is estimated to be}{
We fit a simple formula for $v_{\mathrm{fra}}$ using the values of $v_{\mathrm{fra}}$ for $N_{\mathrm{tar}} =$ 131072 with $N_{\mathrm{tar}} / N_{\mathrm{pro}} \le 4$, 262144 with $N_{\mathrm{tar}} / N_{\mathrm{pro}} \le 64$, and 524288 with $8 \le N_{\mathrm{tar}} / N_{\mathrm{pro}} \le 64$.
The rough fitting formula of these $v_{\mathrm{fra}}$ is then estimated to be
}
\begin{equation}
  v_{\mathrm{fra}} \approx (v_{\mathrm{gl}}^2 + v_{\mathrm{gh}}^2)^{1/2}
  \mathrm{,}
  \label{equ:critical_collisional_growth_velocity_fit}
\end{equation}
where
\begin{equation}
  v_{\mathrm{gl}} = 50 \times \left( \frac{N_{\mathrm{tar}}}{N_{\mathrm{pro}}} \right) ^{-1.3} ~ \mathrm{[m ~ s^{-1}]}
  \mathrm{,}
  \label{equ:critical_collisional_growth_velocity_fit_low}
\end{equation}
and
\begin{equation}
  v_{\mathrm{gh}} = 13 \times \left( \frac{N_{\mathrm{tar}}}{N_{\mathrm{pro}}} \right) ^{0.48} ~ \mathrm{[m ~ s^{-1}]}
  \mathrm{.}
  \label{equ:critical_collisional_growth_velocity_fit_high}
\end{equation}
\added{
The rough fitting formulae of $v_{\mathrm{tra}}$ for $N_{\mathrm{tar}} = 262144$ are estimated to be
\begin{eqnarray}
  v_{\mathrm{tra}} \approx \left \{ \begin{array}{l}
    76 \times (N_{\mathrm{tar}} / N_{\mathrm{pro}})^{-0.67} ~ \mathrm{m ~ s^{-1}} \\
    14 \times (N_{\mathrm{tar}} / N_{\mathrm{pro}})^{1.2} ~ \mathrm{m ~ s^{-1}}
  \end{array} \right . \left . \begin{array}{l}
    \mathrm{for} ~ v_{\mathrm{tra}} \lesssim 40 ~ \mathrm{m ~ s^{-1}} \mathrm{,} \\
    \mathrm{for} ~ v_{\mathrm{tra}} \gtrsim 40 ~ \mathrm{m ~ s^{-1}} \mathrm{,}
  \end{array} \right .
  \label{equ:critical_collisional_transfer_velocity_fit}
\end{eqnarray}
whereas the cases with $N_{\mathrm{tar}} = 262144$ give the largest region of the mass transfer as shown in the right panel of Figure \ref{fig:appendix_Ntop_vcri_Ntar}.
}

\begin{figure}[p]
  \figurenum{14}
  \plotone{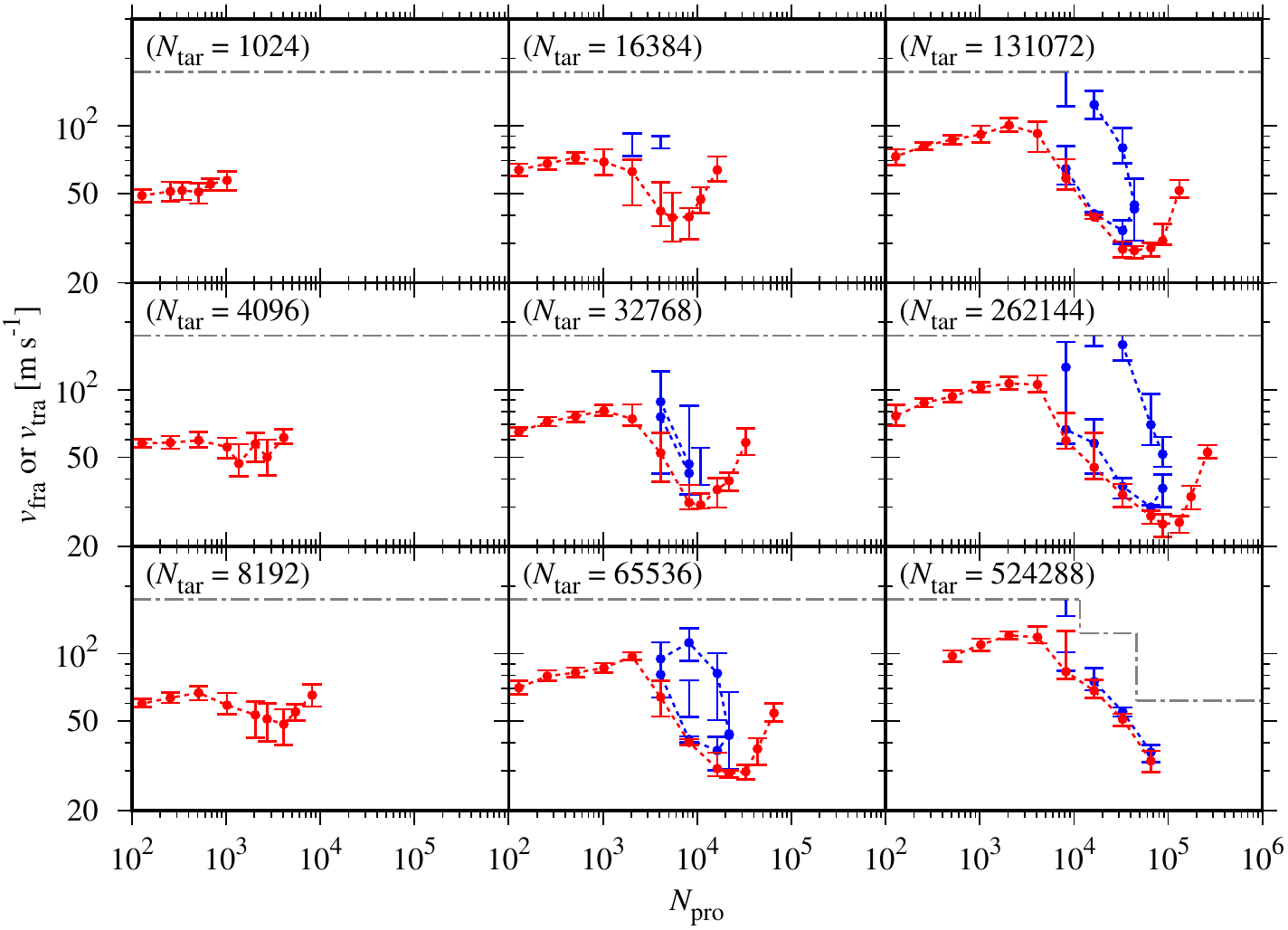}
  \caption{Same as Figure \ref{fig:appendix_Ntop_vcri_Ntar}, but the horizontal axis is the monomer number of the projectile, $N_{\mathrm{pro}}$. Colors represent $v_{\mathrm{fra}}$ (red) or $v_{\mathrm{tra}}$ (blue). The panels are labeled by $N_{\mathrm{tar}}$.}
  \label{fig:appendix_Npro_vcri_Ntar_multi}
\end{figure}

\begin{figure}[p]
  \figurenum{15}
  \plotone{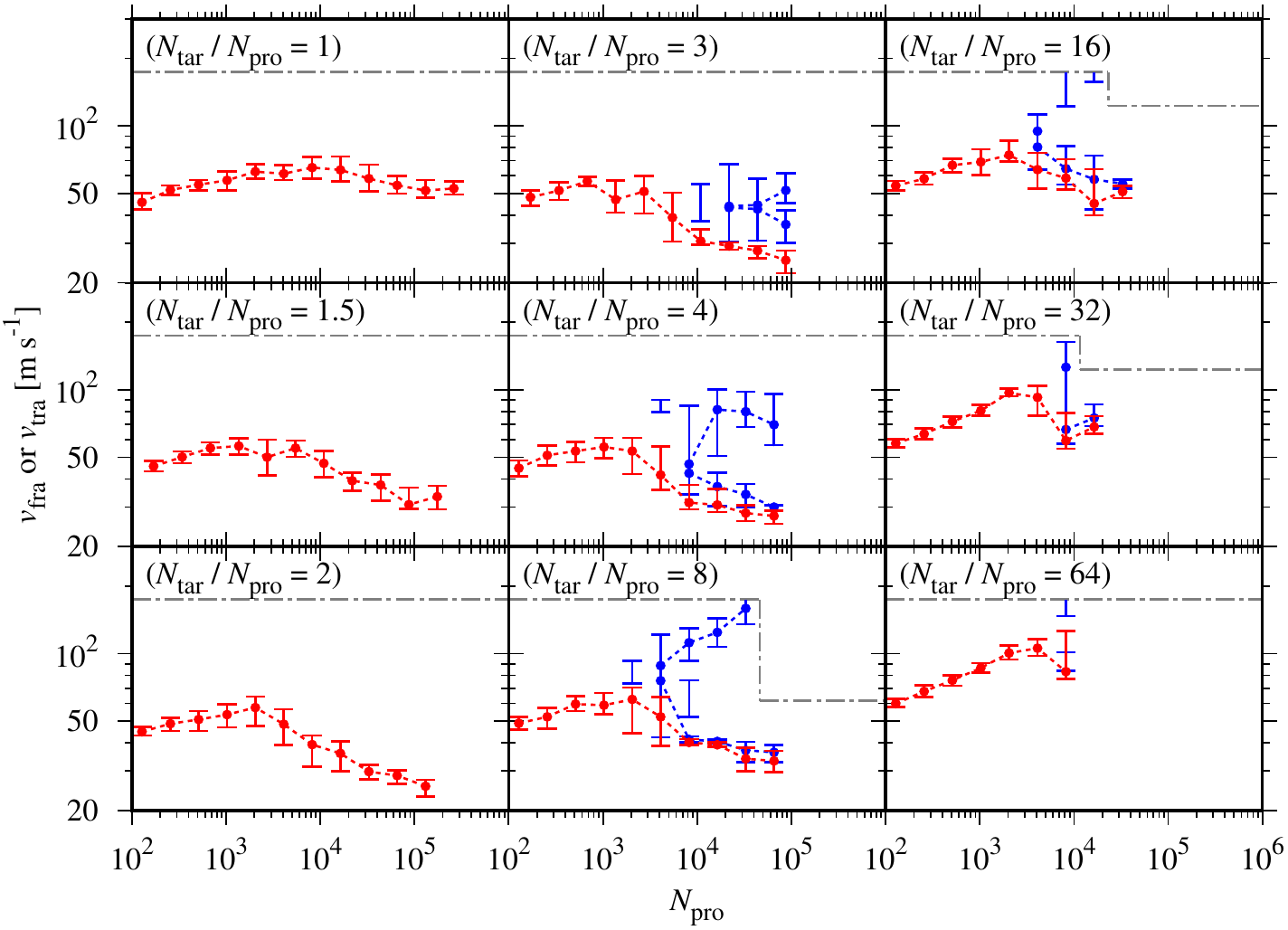}
  \caption{Same as Figure \ref{fig:appendix_Npro_vcri_Ntar_multi}, but the panels are labeled by $N_{\mathrm{tar}} / N_{\mathrm{pro}}$.}
  \label{fig:appendix_Npro_vcri_Ntop_multi}
\end{figure}

Figures \ref{fig:appendix_Npro_vcri_Ntar_multi} and \ref{fig:appendix_Npro_vcri_Ntop_multi} show $v_{\mathrm{fra}}$ and $v_{\mathrm{tra}}$ against $N_{\mathrm{pro}}$ and $N_{\mathrm{tar}}$.
Figure \ref{fig:appendix_Npro_vcri_Ntar_multi} shows that $v_{\mathrm{fra}}$ becomes maximum around $N_{\mathrm{pro}} \approx 3 \times 10^3$ for $N_{\mathrm{tar}} \gtrsim 10^5$, i.e., that $v_{\mathrm{fra}}$ tends to increase with $N_{\mathrm{pro}}$ for the collision between the very large target and the very small projectile, $N_{\mathrm{pro}} \lesssim 10^3$.
This tendency is also shown in Seizinger et al. (2013) and Schr\"{a}pler et al. (2018).

The results with $N_{\mathrm{tar}} = 8192$ and $N_{\mathrm{tar}} / N_{\mathrm{pro}} = 1$ give $v_{\mathrm{fra}} = 65_{-7}^{+7} ~ \mathrm{m ~ s^{-1}}$.
This is in agreement with the critical velocity for $N_{\mathrm{tar}} = 8000$ obtained by Wada et al. (2009).

In collisions with $N_{\mathrm{tar}} / N_{\mathrm{pro}} = 16$, $v_{\mathrm{fra}} = 74_{-5}^{+12} ~ \mathrm{m ~ s^{-1}}$ for $N_{\mathrm{tar}} = 32768$ and $59_{-7}^{+12} ~ \mathrm{m ~ s^{-1}}$ for $N_{\mathrm{tar}} = 131072$.
These are in agreement with the critical values obtained by Wada et al. (2013).
The critical collisional fragmentation velocity for $N_{\mathrm{tar}} = 262144$ and $N_{\mathrm{tar}} / N_{\mathrm{pro}} = 16$ is given by $45_{-5}^{+19} ~ \mathrm{m ~ s^{-1}}$.
This critical value is lower than $\approx 70 ~ \mathrm{m ~ s^{-1}}$, obtained by Wada et al. (2013).
Figure \ref{fig:appendix_Npro_vcri_Ntop_multi} shows that $v_{\mathrm{fra}}$ for $N_{\mathrm{tar}} / N_{\mathrm{pro}} = 16$ tends to decrease with increasing $N_{\mathrm{pro}} \ge 2048$.
Our result also follows the tendency shown in Wada et al. (2013).

For $N_{\mathrm{tar}} / N_{\mathrm{pro}} = 64$, $v_{\mathrm{fra}} = 76_{-4}^{+4}$, $101_{-7}^{+8}$ and $83_{-6}^{+43} ~ \mathrm{m ~ s^{-1}}$ for $N_{\mathrm{tar}} =$ 32768, 131072 and 524288, respectively.
Figure \ref{fig:appendix_Npro_vcri_Ntop_multi} shows that $v_{\mathrm{fra}}$ for $N_{\mathrm{tar}} / N_{\mathrm{pro}} = 64$ tends to increase with $N_{\mathrm{tar}}$.
Our results are in rough agreement with Wada et al. (2013).

\section{Dependence of equal-sized collisions on monomer numbers} \label{sec:app_C}

Results of Wada et al. (2009) suggested that $v_{\mathrm{fra}}$ for $N_{\mathrm{tar}} / N_{\mathrm{pro}} = 1$ increases with monomer numbers of colliding dust aggregates with $N_{\mathrm{tot}} \lesssim 10^4$.
However, our Figure \ref{fig:appendix_Npro_vcri_Ntop_multi} suggests that $v_{\mathrm{fra}}$ tends to decrease with increasing the monomer numbers for $N_{\mathrm{tot}} \gtrsim 10^4$.

\begin{figure}[p]
  \figurenum{16}
  \plotone{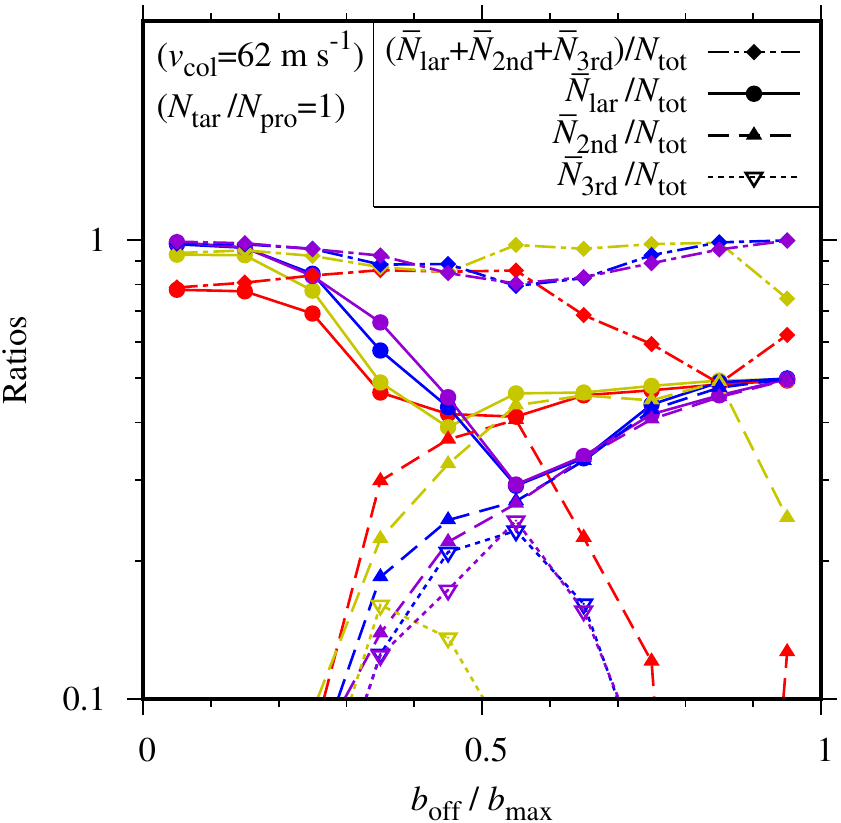}
  \caption{Monomer numbers of the largest, second and third remnants normalized by the total monomer number, ${\bar{N}}_{\mathrm{lar}} / N_{\mathrm{tot}}$, ${\bar{N}}_{\mathrm{2nd}} / N_{\mathrm{tot}}$ and ${\bar{N}}_{\mathrm{3rd}} / N_{\mathrm{tot}}$, and the sum of these remnants, (${\bar{N}}_{\mathrm{lar}} + {\bar{N}}_{\mathrm{2nd}} + {\bar{N}}_{\mathrm{3rd}}) / N_{\mathrm{tot}}$, against the impact parameter $b_{\mathrm{off}} / b_{\mathrm{max}}$ for $N_{\mathrm{tar}} / N_{\mathrm{pro}} = 1$ at $v_{\mathrm{col}} = 62 ~ \mathrm{m ~ s^{-1}}$. Colors are the same as Figure \ref{fig:appendix_vcol_f1f2_Ntar_Ntop_multi_1_2}.}
  \label{fig:appendix_boff_N123_Ntar_vcol62_Ntop1}
\end{figure}

\begin{figure}[p]
  \figurenum{17}
  \plottwo{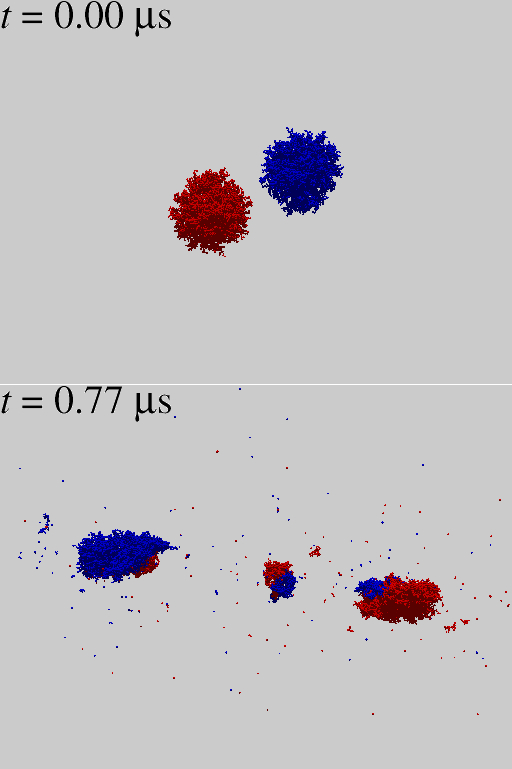}{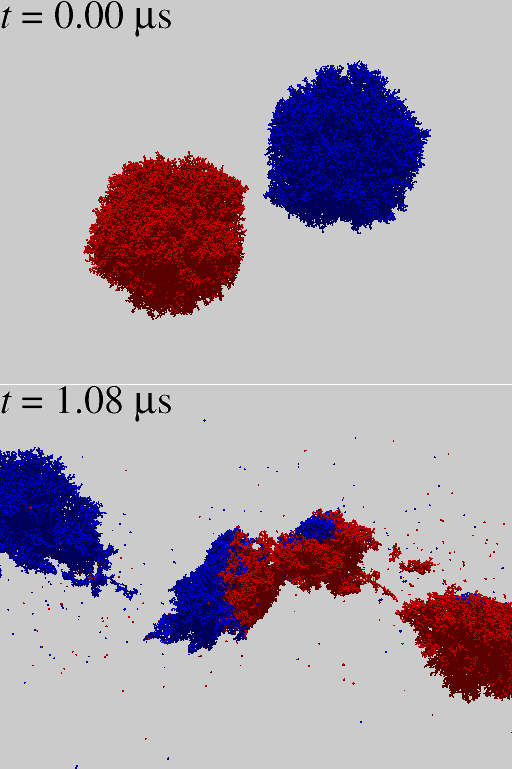}
  \caption{Snapshots of equal-sized offset collisions with $N_{\mathrm{tot}} =$ 16384 (left) and 131072 (right), respectively, and $b_{\mathrm{off}} / b_{\mathrm{max}} = 0.55$ at $v_{\mathrm{col}} = 62 ~ \mathrm{m ~ s^{-1}}$.}
  \label{fig:snapshot_1_55_zx}
\end{figure}

Figure \ref{fig:appendix_boff_N123_Ntar_vcol62_Ntop1} shows the monomer numbers of the largest, second and third remnants normalized by the total monomer number, ${\bar{N}}_{\mathrm{lar}} / N_{\mathrm{tot}}$, ${\bar{N}}_{\mathrm{2nd}} / N_{\mathrm{tot}}$ and ${\bar{N}}_{\mathrm{3rd}} / N_{\mathrm{tot}}$, and the sum of these remnants, (${\bar{N}}_{\mathrm{lar}} + {\bar{N}}_{\mathrm{2nd}} + {\bar{N}}_{\mathrm{3rd}}) / N_{\mathrm{tot}}$, against the impact parameter $b_{\mathrm{off}} / b_{\mathrm{max}}$ and the monomer number of the target $N_{\mathrm{tar}}$ ($= N_{\mathrm{pro}} = N_{\mathrm{tot}} / 2$) for $N_{\mathrm{tar}} / N_{\mathrm{pro}} = 1$ at $v_{\mathrm{col}} = 62 ~ \mathrm{m ~ s^{-1}}$.
Figure \ref{fig:snapshot_1_55_zx} shows examples of equal-sized offset collisions with $N_{\mathrm{tar}} =$ 8192 and 65536.
Our results suggest that collisions with $b_{\mathrm{off}} / b_{\mathrm{max}} \approx$ 0.5-0.7 can make at least three similar-sized remnants that are relatively large but smaller than the target for large $N_{\mathrm{tot}}$, while the number of similar-sized large remnants made in a collision for small $N_{\mathrm{tot}}$ is at most two.
While the maximum monomer number of two similar-sized large remnants is about $N_{\mathrm{tot}}$ / 2, that of three remnants is about $N_{\mathrm{tot}}$ / 3.
It is suggested that this difference between collision outcomes decreases $v_{\mathrm{fra}}$ for collisions of equal-sized dust aggregates with large monomer numbers through decrease of $\langle N_{\mathrm{lar}} \rangle$.

\bibliography{sample63}{}
\bibliographystyle{aasjournal}



\listofchanges

\end{document}